\documentclass[%
aip,
amsmath,amssymb,
reprint,%
]{revtex4-1}

\usepackage{dcolumn}
\usepackage{bm}

\usepackage[utf8]{inputenc}
\usepackage[T1]{fontenc}
\usepackage{mathptmx}

\usepackage{graphicx}
\usepackage{caption}
\usepackage{subcaption}

\usepackage{xcolor}
\usepackage{xfrac}
\usepackage{float}

\begin{document}
	
\title{Effect of inertia on the dynamic contact angle in oscillating menisci}
	
\author{Domenico Fiorini}
\email{domenico.fiorini@keuleuven.be}
\affiliation{%
	von Karman Institute for Fluid Dynamics, Waterloosesteenweg 72, Sint-Genesius-Rode, Belgium
}%
\affiliation{%
	KU Leuven, Dept. of Materials Engineering, Leuven 3001, Belgium
}%
\author{Miguel Alfonso Mendez}
\affiliation{%
	von Karman Institute for Fluid Dynamics, Waterloosesteenweg 72, Sint-Genesius-Rode, Belgium
}%
\author{Alessia Simonini}
\affiliation{%
	von Karman Institute for Fluid Dynamics, Waterloosesteenweg 72, Sint-Genesius-Rode, Belgium
}%
\author{Johan Steelant}
\affiliation{%
	ESTEC-ESA, Keplerlaan 1, Noordwijk, The Netherlands
}%
\affiliation{%
	KU Leuven, Dept. of Mechanical Engineering, Leuven 3001, Belgium
}%
\author{David Seveno}
	\affiliation{%
	KU Leuven, Dept. of Materials Engineering, Leuven 3001, Belgium
}%

\date{\today}
	
\begin{abstract}

The contact angle between a gas-liquid interface and a solid surface is a function of the dynamic conditions of the contact line. Classic steady correlations link the contact angle to the contact line velocity. However, it is not clear whether they hold in presence of inertia and in the case of perfect wetting fluids. We analyze the shape of a liquid interface and the corresponding contact angle in accelerating conditions for two different fluids, i.e. HFE7200 (perfect wetting) and demineralized water. The set-up consists of a U-shaped quasi-capillary tube in which the liquid column oscillates in response to a pressure step on one of the two sides. We obtained the evolution of the interface shape from high-speed back-light visualization, and we fit interface models to the experimental data to estimate the contributions of all the governing forces and the contact angle.\\  Traditional interface models fail to predict the interface shape and its contact angle at large interface and contact line acceleration. We propose a new model to account for the acceleration, and we discuss its impact on the measurement of the transient contact angle.
		
\end{abstract}
	
\maketitle

\section{Introduction}

The curved shape of a gas-liquid interface produces a pressure difference across the interface itself. This is known as capillary pressure and plays a key role in the modeling of the interface motion when the gravity and the capillary forces become comparable \cite{vestad2004flow,liu2017two}. This is the case of many practical applications such as the imbibition of fluids by porous media \cite{ma2012visualization}, enhanced oil recovery operations \cite{kumar2010patterned}, droplet manipulation in microfluidics \cite{willmott2011uptake,willmott2010experimental,liu2017two} and liquid management in space systems \cite{white2019capillary}. In all of these problems the dynamic contact angle, i.e.,
    the angle formed by the interface at a moving contact line, represents the boundary condition for the interface shape and consequently governs the role of the capillary pressure of the system \cite{fernandez2017cohesion}.

The first definitions of dynamic contact angle was given by Blake, Hoffman and Voinov \cite{blake1969kinetics,hoffman1975study,voinov1976hydrodynamics}. These studies show that the motion of the contact point and the contact angle are strongly correlated. In the case of viscous dominated flows \cite{blake1969kinetics,voinov1976hydrodynamics,cox1986dynamics,petrov1992combined,shikhmurzaev1993moving, de1990dynamics,hoffman1975study,tanner1979spreading,jiang1979correlation,bracke1989kinetics,kistler1993hydrodynamics}, the dynamic contact angle can be predicted from the Capillary number $Ca=\mu u_c/ \sigma$, where $\mu$ is the dynamic viscosity of the liquid, $u_c$ is the contact line velocity and $\sigma$ is the surface tension between gas and liquid. Both theoretical and empirical correlations were developed for the case of steady contact line motion. However, several authors \cite{ting_perlin_1995,bian2003liquid,quere1997inertial, shardt2014inertial,willmott2020inertial} have shown that these approaches fail in inertia dominated conditions, and more experimental data is required to clarify the role of contact line acceleration and the history of the contact line dynamics \cite{ting_perlin_1995,bian2003liquid}.

Experimental studies on dynamic contact angles emphasize the difficulty to obtain reliable results due to the microscopic scales involved and thus the high resolution needed for the optical equipment to visualize the interface. Consequently, the results are often affected by high uncertainty and generalize poorly. \citet{petrov1993quasi} consider the dynamic contact angle as "inaccessible" by optical means and attempt to overcome this limit by computing the contact angle from the theoretical quasi-static interface shape fitted to the experimental data. This theoretical interface shape is obtained by solving a boundary value problem which \citet{petrov1993quasi} closed using Neumann boundary conditions, i.e. the interface slope on the wall and far from the solid surface. 

Methods for measuring the contact angle using an analytic formulation of the meniscus interface are referred to as Meniscus Profile Methods (MPM)  \cite{iliev2011dynamic}. \citet{maleki2007landau} extended the approach of  \citet{petrov1993quasi} to a "quasi-steady" approximation where the shape of the gas-liquid interface depends on the capillary number. \citet{iliev2011dynamic}  uses the same approximation, closing the boundary value problem using the interface locations near the wall, available experimentally. In the case of receding interfaces, \citet{maleki2007landau} found a critical capillary number above which the contact angle seemingly jumps to zero and a liquid film is deposed along the walls. Voinov and Eggers \cite{voinov2000wetting,eggers2005existence} also hypothesized the existence of such condition. 

The case of inertia-dominated interfaces has not been considered in the framework of MPM methods. Inertia can have an important role in classic problems such as the capillary rise, wherein inertia and surface tension control the initial accelerating phase. The typical modelling approach for this problem (see \citet{quere1997inertial}) considers a quasi-steady formulation for the contact angle and assumes that the interface remains spherical also in dynamic conditions. These models have been proved valid for high viscous fluids over a large range of Ca (see \cite{quere1997inertial,Wu2017}) and in microgravity conditions \cite{stange2003capillary}. These have also been used to describe the results from experiments in short tubes \citep{shardt2014inertial}, i.e. for which liquid can spread onto the external walls after reaching the top of the tube, and the inertial uptake of liquid drops \cite{willmott2020inertial}, for which the capillary suction due to the meniscus interacts with the Laplace pressure in the (shrinking) drop.

However, as also reported by \citet{quere1997inertial}, these models are incapable of describing capillary rise experiments with less viscous fluids, especially in the presence of oscillations of the liquid column. Whether the discrepancy is due to the incorrect modelling of the interface shape, the miss-prediction of the contact angle evolution, the modelling of viscous forces or the modeling of the pressure drop at the tube entrance remains an open question.

In this work, we analyze two of the aforementioned modelling challenges, namely the description of the gas-liquid interface and the contact angle dynamics. The chosen configuration is a U-tube geometry, where an axial-symmetric gas-liquid interface forms on the two sides. The tube radius was selected to maximize the impact of capillary forces on the interface force balance. We measure the dynamic contact angle using an adapted MPM and introduce an empirical term to account for the impact of inertia in circular tubes. We analyze the interface dynamics for two low viscous fluids, namely demineralized water and HFE 7200 by 3M Novec.

The experimental configuration is described in section \ref{sec:experimental_methods}.  The interface parametrization is described in section \ref{sec:modeling}. Results are presented in section \ref{sec:results} while section \ref{sec:conclusion} closes with conclusions and outlooks to future works.


\section{Experimental methods}\label{sec:experimental_methods}
Table \ref{tab:properties} shows the relevant properties for the fluids considered in this study, namely demineralized water and pure HFE7200 ($\text{C}_\text{4} \text{F}_\text{9} \text{O} \text{C}_\text{2} \text{H}_\text{5}$) produced by 3M Novec. The HFE7200 is a synthetic liquid in the family of hydrofluoroethers characterized by high volatility (vapor pressure 14 kPa at $25^{\circ}C$) and small equilibrium contact angles (below 10 degrees). These liquids are frequently used to simulate the properties of cryogenic space fluids \cite{rausch2015density}. 

\begin{table}[h]
	\caption{\label{tab:properties} Fluids physical properties at 298 K}
	\begin{ruledtabular}
		\begin{tabular}{cccc}
			 &   & Water & HFE7200\\ \\
			\hline
			density $(\rho)$ & $\mbox{kg/m}^3$ & 996.66 & 1423.00 \\
            dynamic viscosity $(\mu)$ & $\mbox{mPa}\cdot \mbox{s}$ & 0.97 & 0.64 \\
            surface tension $(\sigma)$ & $\mbox{mN/m}$ & 72.01 & 13.62 \\
		\end{tabular}
	\end{ruledtabular}
\end{table}

\begin{figure}[h]
\centering
\includegraphics[width=20pc]{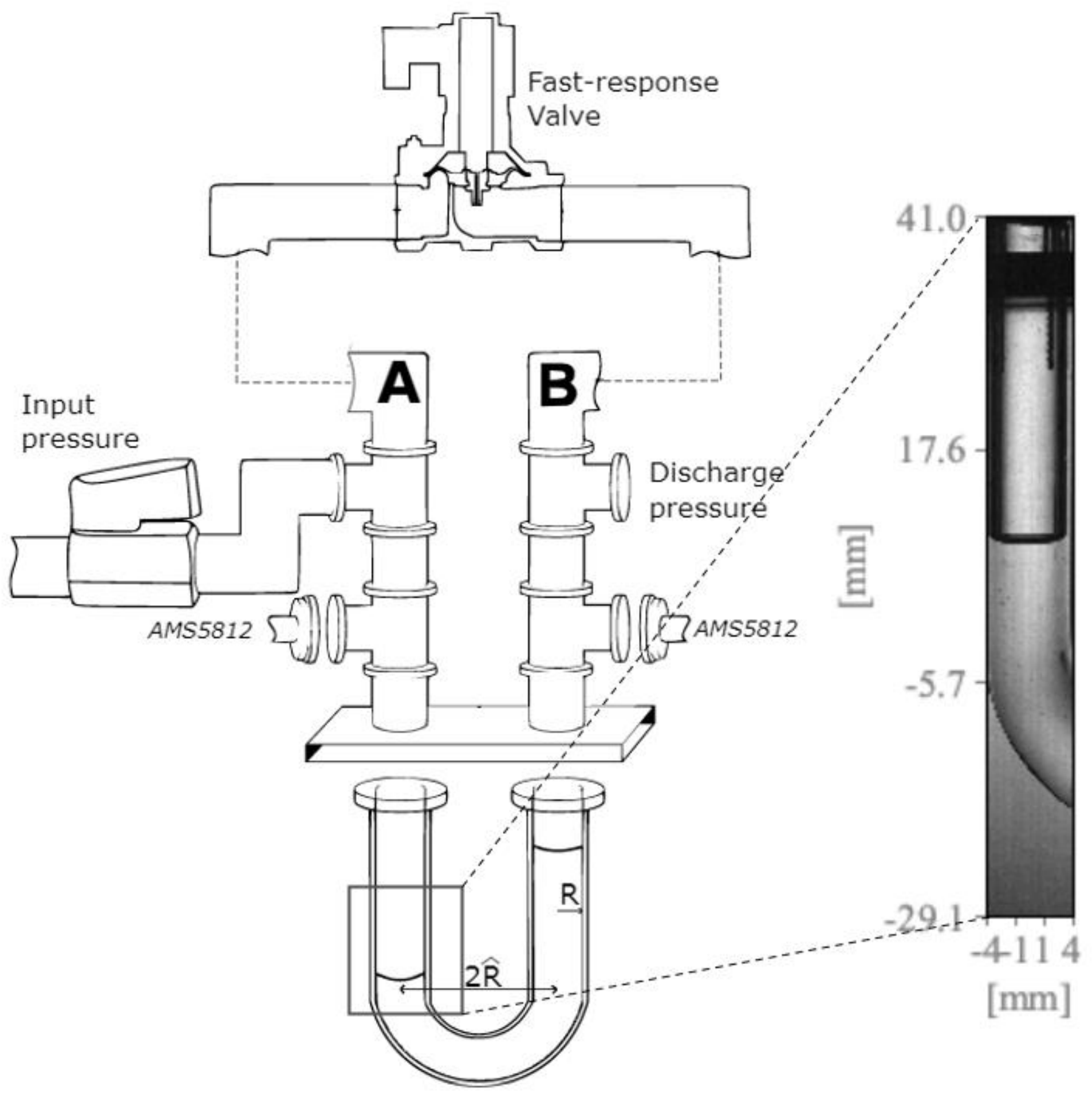}%
\caption{\label{fig:setup} Left: schematic diagram of experimental setup. At initial conditions, side A is pressurized and side B is at room pressure, a fast response valve allows for suddenly releasing the pressure difference to begin the dynamic experiment. Right: example of complete image acquired.}
\end{figure}

The behavior of the gas-liquid interface of HFE7200 and water is analyzed in a U-shaped quartz tube manufactured by PIERREGLAS (Belgium). The U-tube internal diameter is 8 mm with axial length 164 mm (see Figure \ref{fig:setup}). The two sides of the tube (A and B in Figure \ref{fig:setup}) are connected by flexible pipes. A fast response valve controls the connection between these sections. Two differential piezo-resistive measuring cells (AMS 5812, ANALOG MICROELECTRONICS) are placed at the two entrances of the tube. The Side A is connected to an input pressure line, while the side B is connected to a discharge pressure line. The first experimental campaign was carried out with HFE7200.

The tube was kept as provided protected from the external environment until the test. Then, the tube internal surface was rinsed with an isopropanol solution and let dry before closing all the connections with the laboratory environment. After the tests with HFE7200, the tube was emptied, dried at ambient conditions, and prepared for the test with water using the same procedure.

Initially, the U-tube is filled with the test liquid while the fast response valve connecting the two sides is open, and the two interfaces of the liquid column have the same height. The initial conditions of the experiment are defined by closing the fast response valve and increasing the pressure on side A. This sets the initial level difference between the two interfaces and zero velocity of the interface. 

The preparatory phase ends by closing all connections to the external environment and waiting sufficient time for HFE7200 to saturate the gas volumes within the tube. The experiment starts by opening the fast-response valve, and the two liquid interfaces start oscillating around the equilibrium position. The motion of one of the two interfaces is recorded with a high-speed camera SP-12000-CXP4 (STEMMER IMAGING), acquiring grey-scale images at 300 fps. The interface shape is visualized by a diffused light source on the opposite side of the tube. The active region of the camera sensor is restricted to the central portion with 4096x512 pixels to achieve the highest acquisition frequency allowed by the camera. The camera mounts a Nikon micro-Nikkor 105mm lens and is positioned to visualize the motion of the interface while spanning the full tube width, as shown in Figure \ref{fig:setup}. Each frame is analyzed with image processing techniques and the interface shape is converted in a set of points $(r_i, h_i )$ with $i\in[1,n_p]$ and $n_p$ the number of points detected. Details about the image processing and the correction for optical distortions are given in the supporting information.

\subsection{Image analysis}\label{subsec:image_analysis}
\begin{figure}[h]
\centering
\begin{minipage}{6pc}
\subcaptionbox{\footnotesize{Original.}\label{subfig:a}}
{\includegraphics[width=6pc, trim = 5 5 8 5, clip]{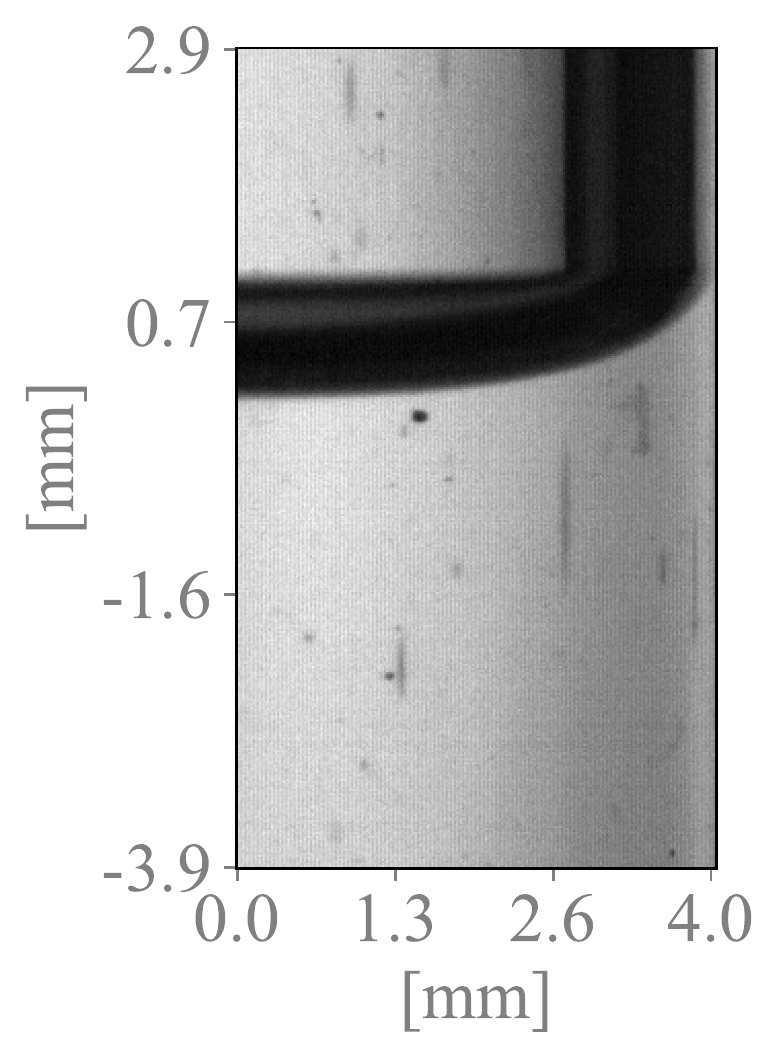}}
\end{minipage}\hspace{0pc}%
\begin{minipage}{6pc}
\subcaptionbox{\footnotesize{Recontrasted.}\label{subfig:b}}
{\includegraphics[width=6pc, trim = 5 5 8 5, clip]{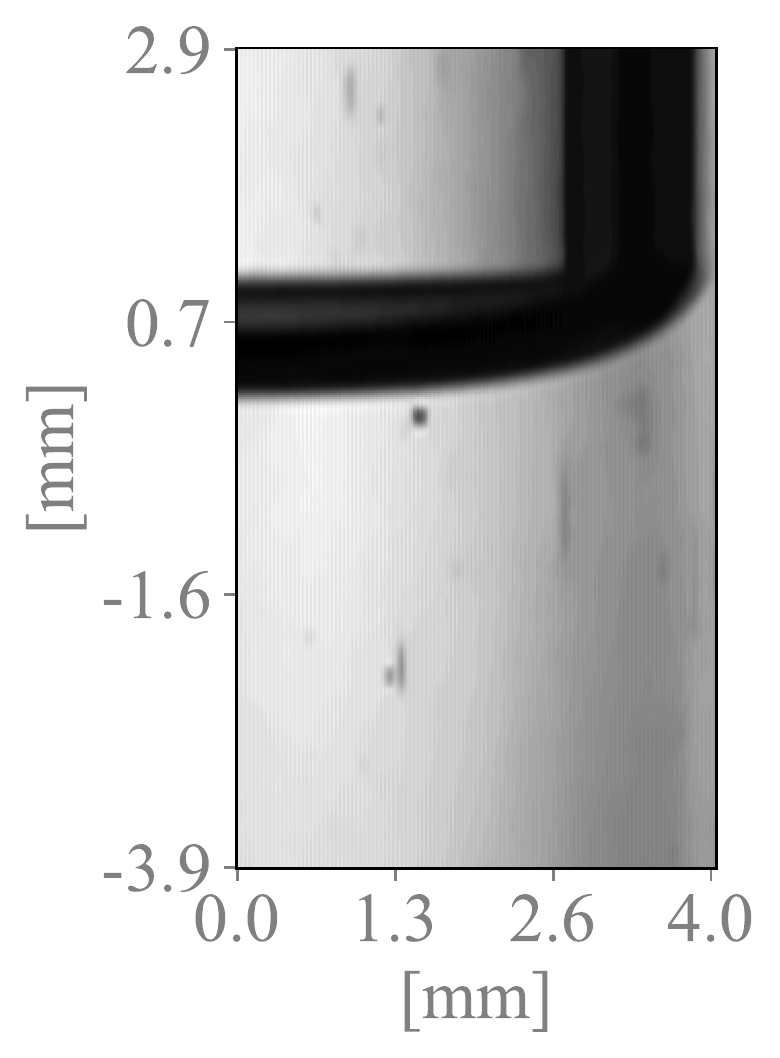}}
\end{minipage}\hspace{0pc}%
\begin{minipage}{6pc}
\subcaptionbox{\footnotesize{Gradients.}\label{subfig:c}}
{\includegraphics[width=6pc, trim = 5 5 8 5, clip]{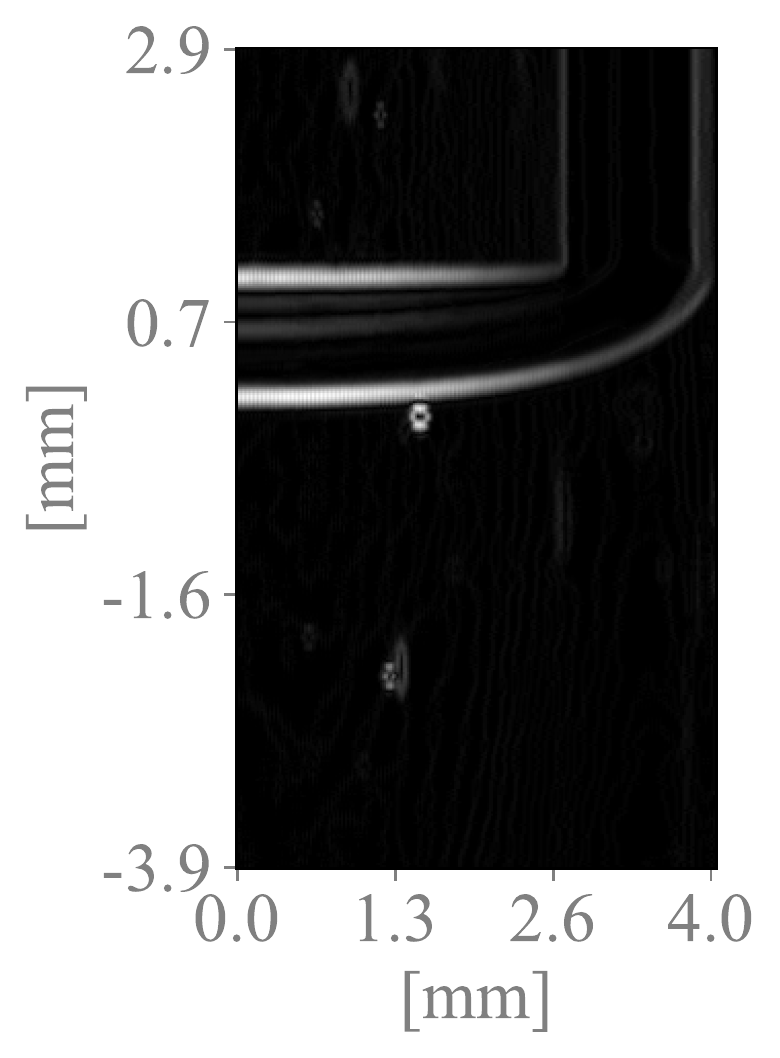}}
\end{minipage}\hspace{0pc}%
\begin{minipage}{6pc}
\subcaptionbox{\footnotesize{Edges.}\label{subfig:d}}
{\includegraphics[width=6pc, trim = 5 5 8 5, clip]{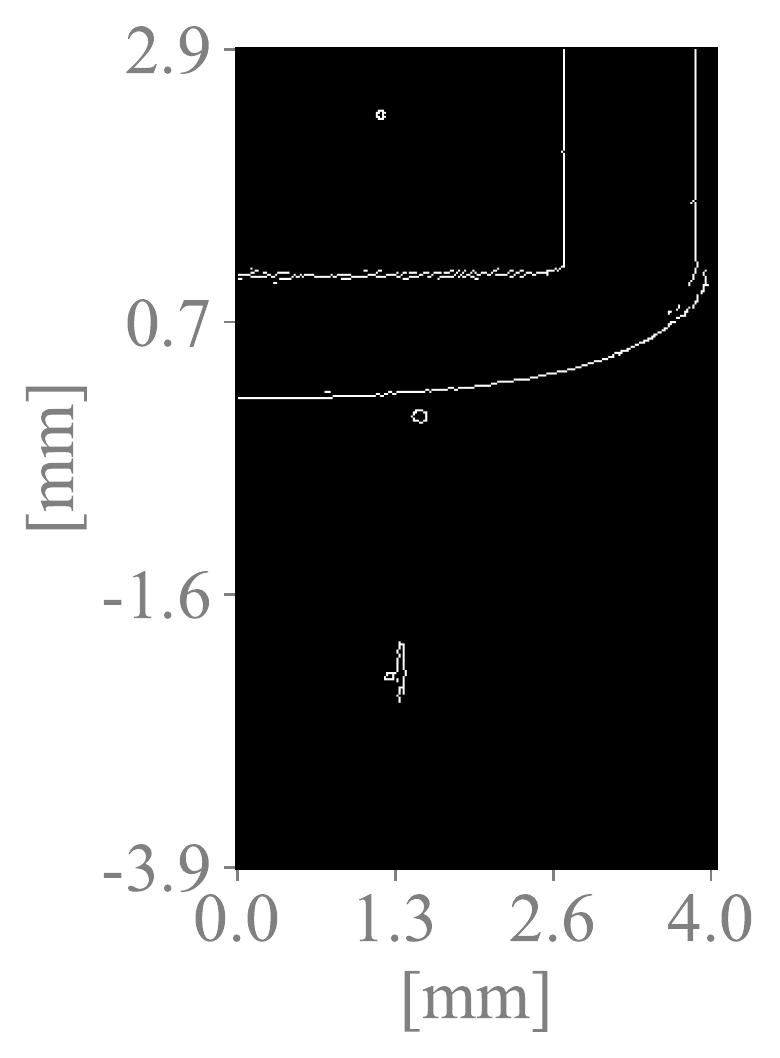}}
\end{minipage}\hspace{0pc}%
\begin{minipage}{6pc}
\subcaptionbox{\footnotesize{Interface.}\label{subfig:e}}
{\includegraphics[width=6pc, trim = 5 5 8 5, clip]{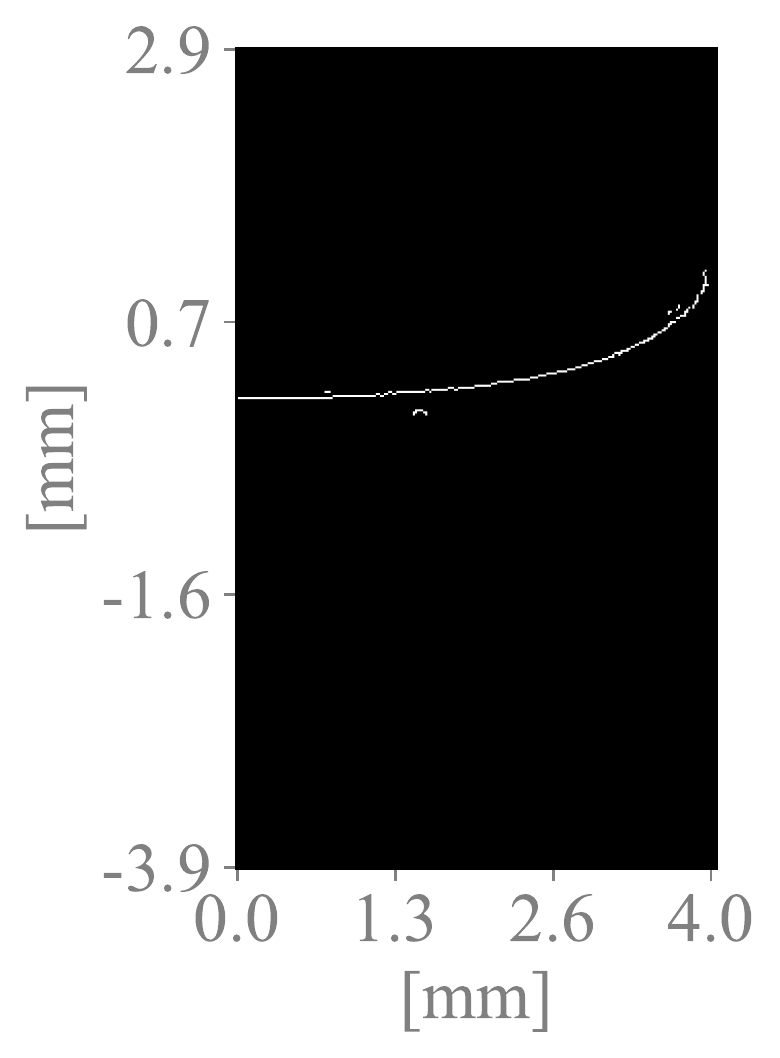}}
\end{minipage}\hspace{0pc}%
\begin{minipage}{6pc}
\subcaptionbox{\footnotesize{Result.}\label{subfig:f}}
{\includegraphics[width=6pc, trim = 5 5 8 5, clip]{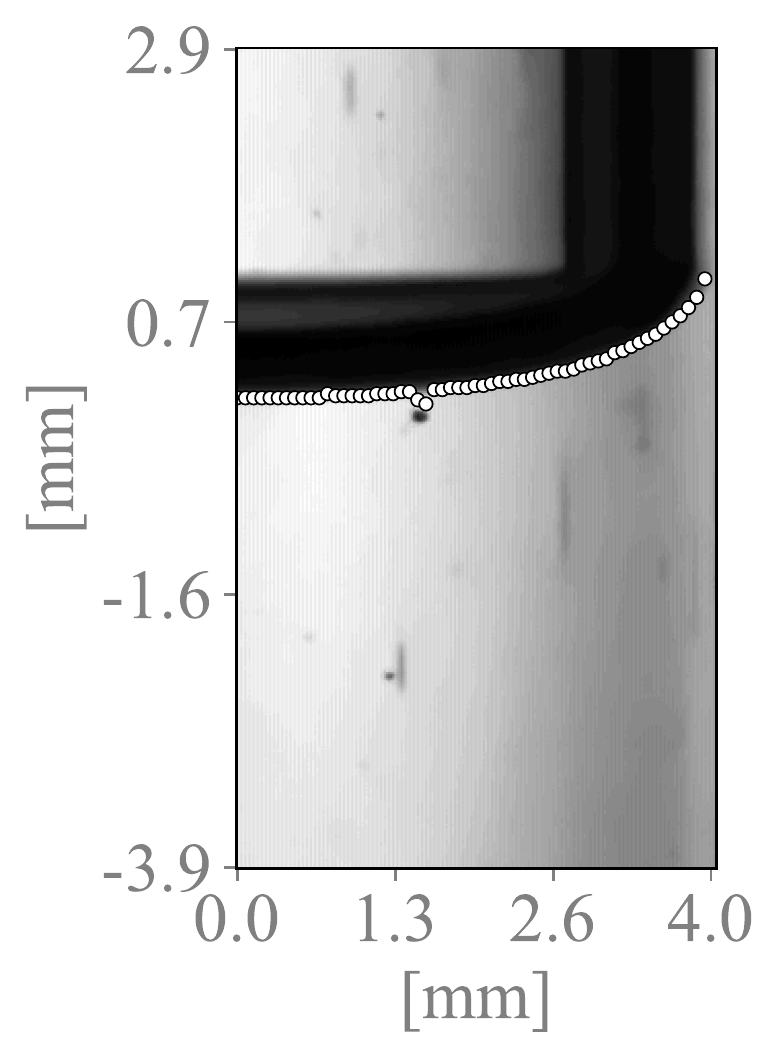}}
\end{minipage}\hspace{0pc}%
\caption{\label{fig:edge_detection} Sequence of image processing steps implemented to identify the liquid interface}
\end{figure}

 The image processing for the interface detection consists of four main steps, recalled in Figure \ref{fig:edge_detection}. In the first step, the original images (Figure \ref{subfig:a}) are filtered via global noise removal based on n-local means denoising algorithm \cite{buades2011non}, followed by a first image recontrasting (Figure \ref{subfig:b}). The second and third steps process each column of the image separately. The grayscale profiles are first filtered using a Savitzky–Golay filter \cite{savitzky1964smoothing} and the resulting image re-contrasted using directional grayscale averaging \cite{mendez2016measurement}. This allows for increasing the grey level intensity in the region of the liquid interface close to the tube walls, which appear darker compared to the centre of the channel.

\subsection{Optical correction}\label{subsec:optical_correction}

\begin{figure}[h!]
    \centering
     \hspace*{-0.5cm} 
    \includegraphics[width=9cm,trim={0cm 1.5cm 0cm 1.5cm},  clip]{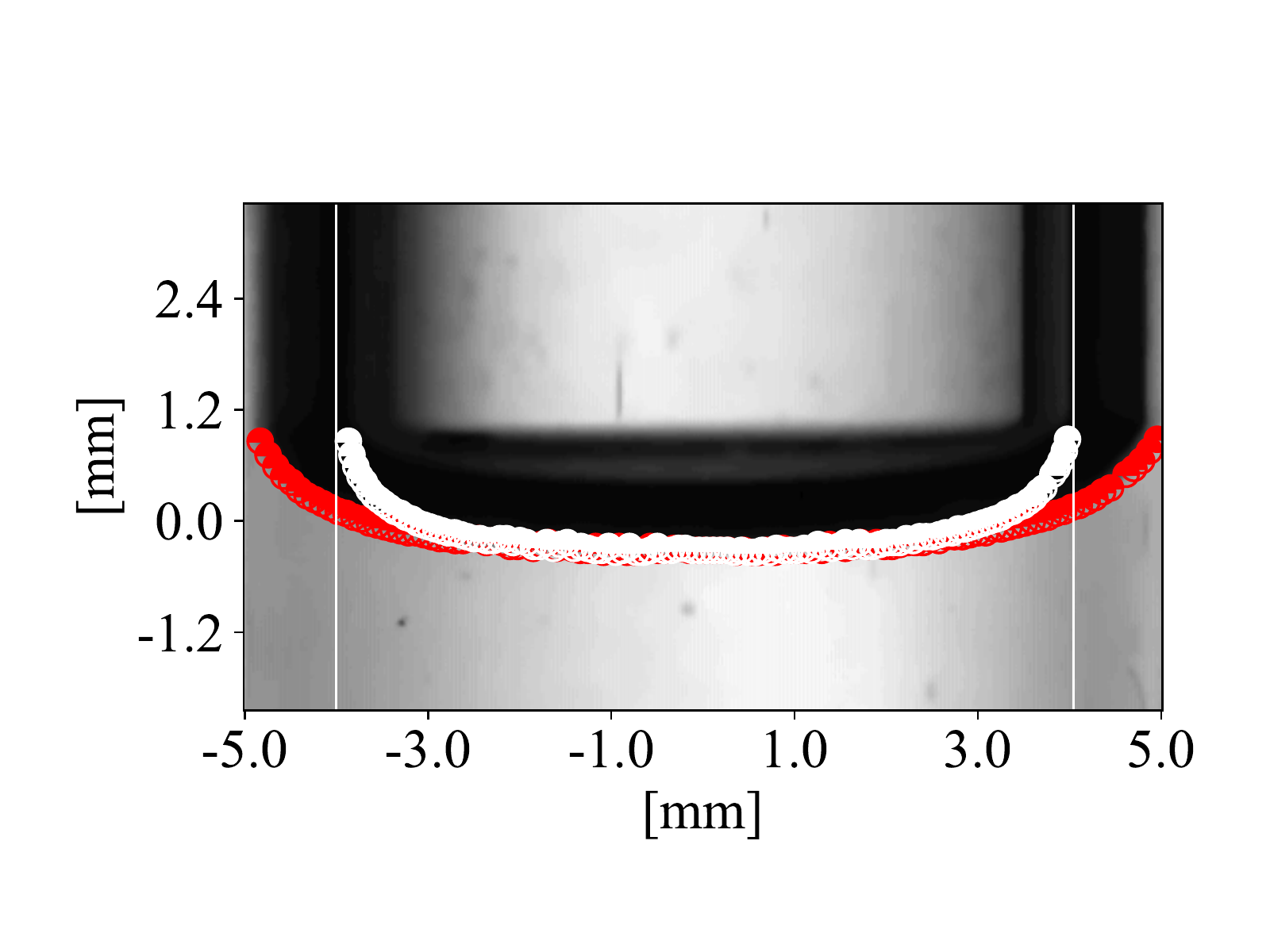}
    \caption{Example image of HFE7200 interface without correction (red circles) and with correction (white circles). The vertical lines corresponds to the real inner surface of the tube.}
    \label{fig:opticalCorrection}
\end{figure}

\begin{figure}[h!]
    \centering
     \hspace*{-0.1cm} 
    \includegraphics[width=7.5cm,trim={0cm 0cm 0cm 0.5cm},  clip]{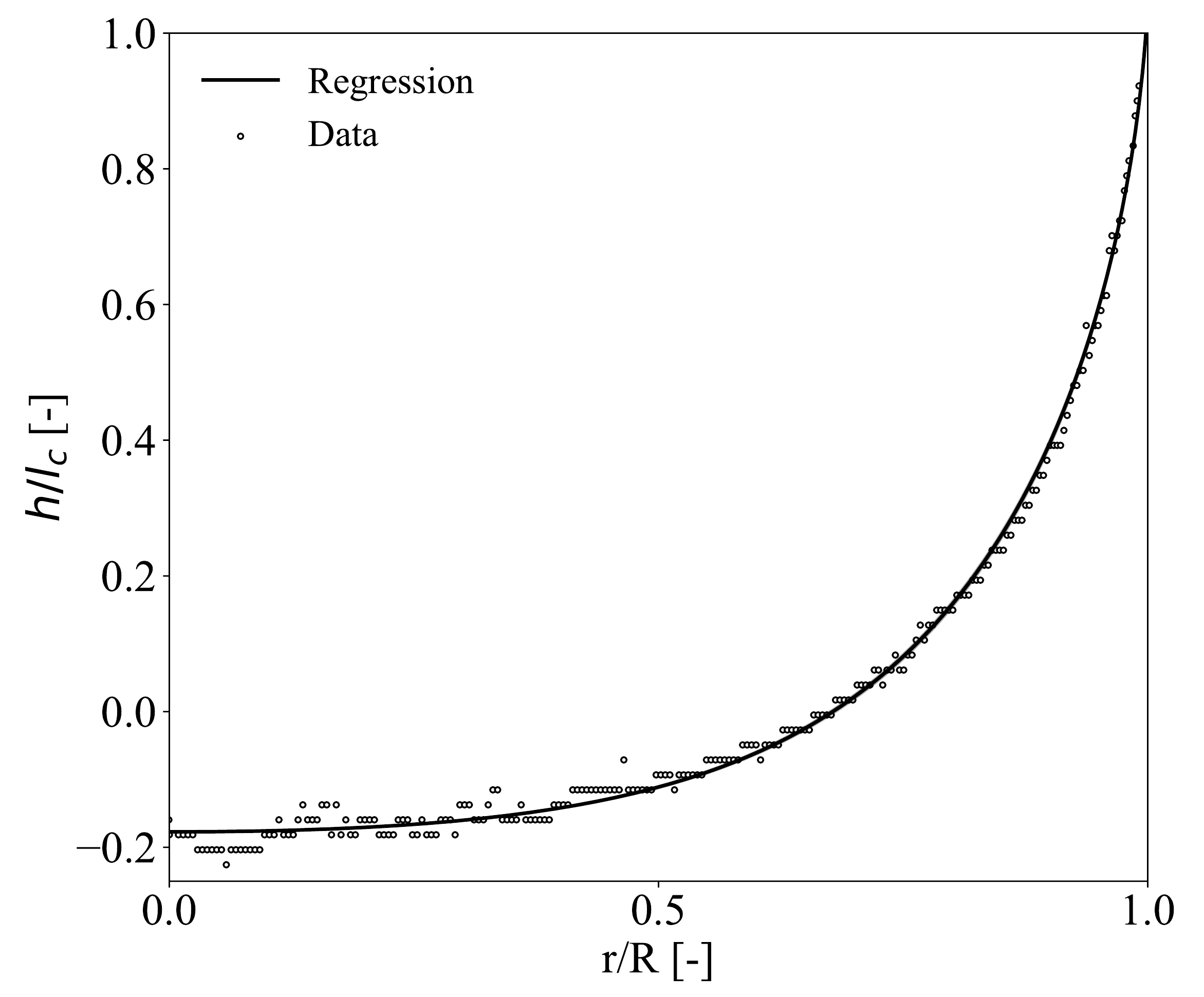}
    \caption{Corrected HFE7200 interface profiles and regression with static interface with $l_c = 0.99 mm$ and $\theta_S = 8.6$.}
    \label{fig:HFEstaticRegression}
\end{figure}

\begin{figure}[h!]
    \centering
    \includegraphics[width=9cm]{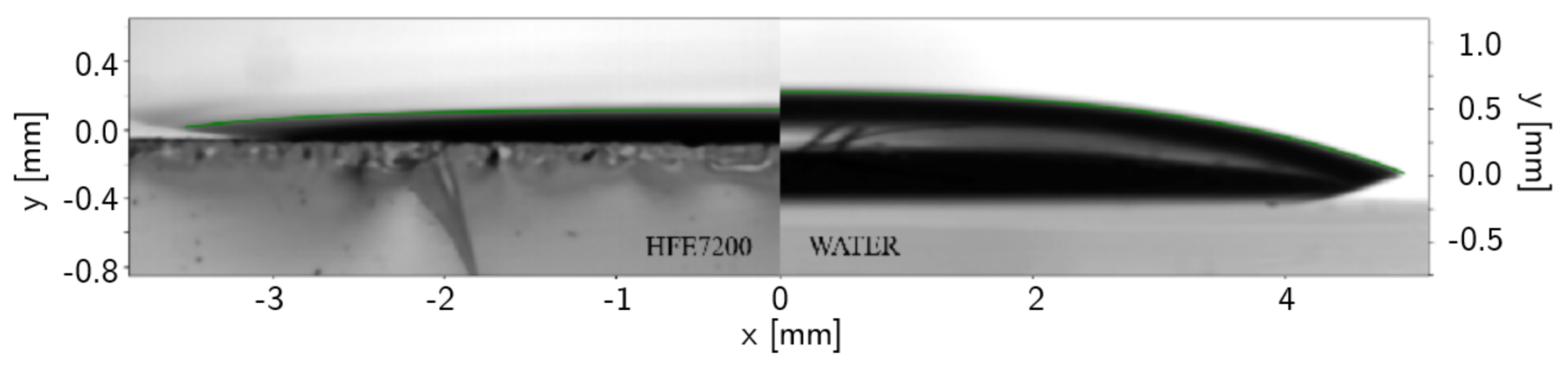}
    \caption{Side view of HFE7200 droplet (left) and water droplet (right) for static contact angle quantification. x =0 coincides with the apex of the droplet while y = 0 corresponds to the positioning of the contact point within the image. The y-axis has a different scale on the left and right side.}
    \label{fig:sessileDroplet}
\end{figure}

The circular shape of the tube deforms the image of the meniscus in the radial direction. This implies that the scaling of the image follows a non linear trend and the size of one pixel in the radial direction varies from $21 \ \mu m$ in the channel center to $9 \ \mu m$ in the channel wall. The size of one pixel is $21 \ \mu m$ along the channel axis. The meniscus appears larger than its actual size, and the contact point with the solid surface seemingly lies in the region between the inner and outer side of the tube, as shown by the red points in Figure \ref{fig:opticalCorrection}. The distortion is maximum at the contact point of the meniscus and disappears at the centre of the channel. Following \citet{darzi2017optical}, the distortion can be corrected given the distance of the camera from the tube center and the refraction index of the materials involved.
We validate the procedure by comparing the corrected profiles with the theoretical profile in static conditions (see Eq. \ref{eq:quasi-static} in section \ref{subsec:inertialless}). A similar approach is also used by \citet{petrov1993quasi}. In static conditions, the interface shape is determined by the static contact angle $\theta_S$ and the capillary length $l_c=\sqrt{\sigma/{\rho g}}$. 

The regression of the static profiles with the static meniscus model (see Figure \ref{fig:HFEstaticRegression}) yields $l_c = 0.98 \pm 0.01 \ mm$ and $\theta_s = 6.57^{\circ} \pm 1.48^{\circ}$ for HFE7200, $l_c = 2.72 \pm 0.1 $ mm and $\theta_s = 22.4^{\circ} \pm 1.6^{\circ}$ for demineralized water. The capillary length computed from the liquid properties at the laboratory temperature ($25^{\circ}$, see Table \ref{tab:properties}) is $l_c=0.98$ mm for HFE7200 and $l_c=2.71$ mm for demineralized water. Moreover, the static contact angle obtained from the side view of HFE7200 and water droplets (see Figure \ref{fig:sessileDroplet}) deposited on a flat substrate made of an equally prepared quartz material yields $\theta_s=7.1^{\circ} \pm 2^{\circ}$ for HFE7200 and $\theta_s=20.33^{\circ} \pm 1^{\circ} $ for demineralized water. This validates both the static model formulation and the optical correction implemented in this work.

As expected, the corrected profile (white circles in Figure \ref{fig:opticalCorrection}) lies inside the tube and wets its inner surface (vertical white lines in Figure \ref{fig:opticalCorrection}).


\section{Contact Angle Measurement}\label{sec:modeling}

Both water and HFE7200 wet the inner walls of the quartz tube. Several authors have shown that the simple Tangent-Line methods (TLM), i.e. a linear extrapolation of the gas-liquid interface to the wall, tends to overestimate the contact angle \cite{petrov1993quasi,maleki2007landau,iliev2011dynamic}, especially with perfect wetting liquids like HFE7200.

The MPM used in this work retrieves the contact angle by fitting a model for the interface to the available data. This model depends on various parameters, among which the contact angle. The contact angle measure is thus cast as a regression problem and the accuracy of the method depends on the accuracy of the interface model. We review three interface models in this section. The first two are the static and quasi-steady models (\ref{subsec:inertialless}), used by various authors \cite{petrov1993quasi,maleki2007landau,iliev2011dynamic, snoeijer2013moving}. We then move to a new approach to account for the inertia contribution in subsection \ref{subsec:inertia}.


\begin{figure}
    \centering
    \includegraphics[width=7.5cm, trim = 10 30 10 10, clip]{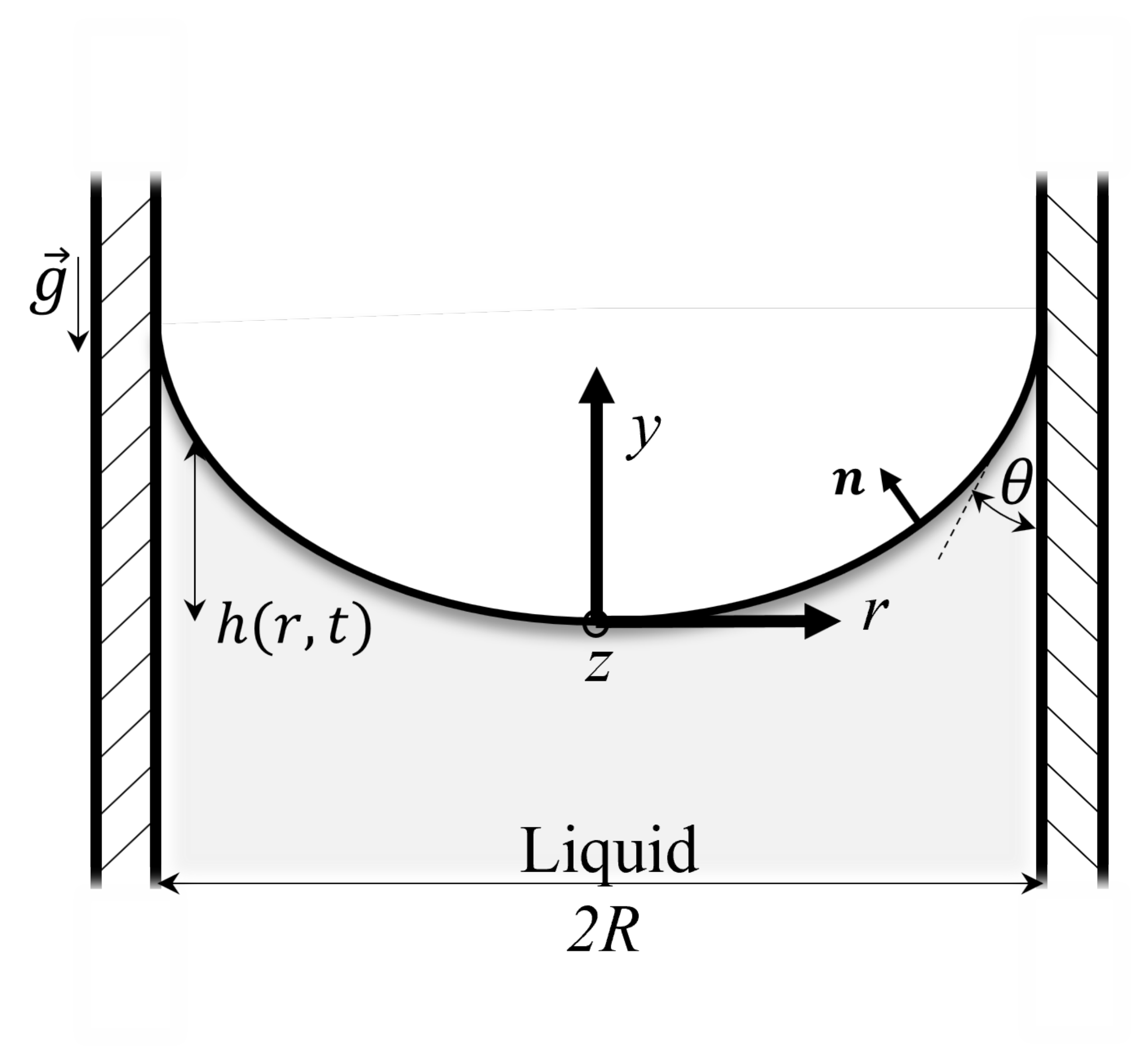}
    \caption{Schematics of the interface model.   The origin is located at the  interface,  in  the  center  of  the  channel,  and  moves  with  it,  i.e. h(0,t) = 0 for all times.}
    \label{fig:sketch}
\end{figure}

\subsection{Inertial-less modeling}\label{subsec:inertialless}

Figure \ref{fig:sketch} illustrates the definition of the interface model and the relevant parameters. At rest, the shape is controlled by the balance of gravitational and capillary forces \cite{landau1959fluid}. The resulting interface is axial symmetric, and the force balance in an infinitesimally thin control volume bounded by the interface at $y=h(r)$ and $y=0$ reads:
\begin{equation}
      \sigma \nabla \cdot \mathbf{n}+\rho g h =0\\\label{eq:staticBalance}
\end{equation}
where $\mathbf{n}$ is the unit normal vector and $\nabla\cdot \mathbf{n}$ is the interface curvature. For an axial-symmetric meniscus, these are: 
\begin{equation}
    \label{eq:curvature_def}
    \mathbf{n}=\frac{\mathbf{z}-h_r \mathbf{r}}{1+h^2_r} \quad \quad \nabla \cdot \mathbf{n}=\frac{-r h_r - r^2 h_{rr}}{r^2 (1+h^2_r)^{3/2}}\,,
\end{equation} having used the short hand notation $h_r$ and $h_{rr}$ to denote the first and second partial derivatives of $h$ with respect to the radial coordinate $r$. Eq. \ref{eq:staticBalance} was solved with Neumann conditions at the wall and Dirichlet conditions at the channel centre. Thus the problem reads:
\begin{equation}
    \label{eq:quasi-static}
\left\{\begin{array}{@{}l@{}}
    \nabla \cdot \mathbf{n}+l^{-2}_c\,h=0\\
    h_r(R,t)=\mbox{ctg}(\theta (t))\\
    h_r(0,t)=0\\
  \end{array}\right.\,, 
\end{equation} having introduced the capillary length $l_c$, and denoting as $\theta(t)$ the dynamic contact angle, defined as the "apparent dynamic contact angle" by \citet{petrov1993quasi}. Model (\ref{eq:quasi-static}) provides a good approximation of the interface shape also for small contact line velocities, and it is often called quasi-static \cite{petrov1993quasi}. In these conditions, the viscous bending of the interface near the contact line is not visible at a `macroscopic' scale. i.e. at the scale of the "apparent" contact angle \cite{snoeijer2013moving}. 

A simple analytic solution can be derived under the assumption that $h_r\ll1$ and thus $\nabla \cdot \mathbf{n}\approx h_{rr}$. In what follows, we do not rely on this simplification and solve \eqref{eq:quasi-static} numerically. It is worth noticing that this model has a singularity at $r=0$, as the curvature approaches zero. This was circumvented by excluding the origin in the computational domain.
 
When the capillary number is not negligible, the viscous forces inside the meniscus must be included in the force balance \cite{maleki2007landau,iliev2011dynamic}. At a constant contact line velocity $u_c$, the viscous pressure can be modeled with the Voinov approximation \cite{voinov1976hydrodynamics}:
\begin{equation}
    \Delta p_\nu = 3\frac{\mu U_{cl}}{(R-r)} F(\beta)\label{eq:voinovApprox}\,,
\end{equation} where $F(\beta)$ is a correction factor \cite{delon2008relaxation} for small slopes of the interface, and $\beta=\cot (h_r)$ is the interface slope. This term can be expressed as $F(\beta)=\sin(\beta(r))/(\beta(r)^2)$. In case of receding contact lines, this approximation holds until a critical capillary number \cite{voinov2000wetting,eggers2005contact}, above which the contact line vanishes and it is impossible to define a contact angle. The `quasi-steady' model is thus a correction of \eqref{eq:quasi-static} to include \eqref{eq:voinovApprox}:
\begin{equation}
    \label{eq:quasi-steady}
\left\{\begin{array}{@{}l@{}}
   \nabla \cdot \mathbf{n}+l^{-2}_c\,h-3\frac{Ca}{(R-r)} F(\beta)=0\\
    h_r(R,t)=\mbox{ctg}(\theta (R,t))\\
    h_r(0,t)=0\\
  \end{array}\right.\,.
\end{equation}

This model is valid for quasi-steady (i.e. with moderate accelerations) motion of the contact line, where the receding capillary number is below the critical value, and the flow field in the meniscus is dominated by pressure and viscous forces (lubrication approximation). 

Because \eqref{eq:quasi-steady} has also a singular point at the wall, the numerical domain does not reach $r=R$, but stops at $r=R-l_d$. The exact value of $l_d$ is under debate, however most authors \cite{voinov1976hydrodynamics, delon2008relaxation, maleki2007landau} place it in the range $10^{-5}-10^{-6}$ mm.


\subsection{Correction for inertia effects}\label{subsec:inertia}
Although the lubrication approximation is valid in the vicinity of the contact line for a large range of conditions, at some distances from walls, the flow inertia impacts the meniscus shape. Given the difficulties in modelling this effect from first principles, we account for its impact with an additional (heuristic) pressure term of the form:
\begin{equation}
   H_a(r,t)= \rho a_i(t) l_h \left(1-e^{-\frac{r-R}{l_i}}\right) \label{eq:pinertia}\,,
\end{equation} where $a_i (t)$ is the acceleration of the interface as a function of time, $l_h= R c_t$ is a characteristic length defined by the model parameter $c_t$, and $l_i (t)$ controls how the contribution of inertia decays towards the tube wall. 
This terms plays the role of a dynamic pressure, arising as the accelerating liquid column adapts to the interface velocity within a certain distance from it. Its mathematical form has been chosen to allow a wide variety of interface shapes depending on the closing parameter $l_i (t)$, to be identified via regression of the interface shape.

For small interface accelerations, $H_a$ is negligible and the quasi-steady form of the equation is retrieved. Summarizing, the inertia-corrected  model has the form:
\begin{equation}
    \label{eq:dynamic}
\left\{\begin{array}{@{}l@{}}
   \nabla \cdot \mathbf{n}+l^{-2}_c\,h-3\frac{Ca}{(R-r)} F(\beta)+H_a(r,t)=0\\
    h_r(R,t)=\mbox{ctg}(\theta (R,t))\\
    h_r(0,t)=0\\
  \end{array}\right.\,.
\end{equation}

In section \ref{sec:results}, we refer to system \eqref{eq:quasi-static} as gravity-based (GB) model, to system \eqref{eq:quasi-steady} as viscous-gravity (VG) model, and to system \ref{eq:dynamic} as Inertia-corrected (IC) model.

\subsection{Interface regression}\label{subsec:inverse_method}

An optimization routine is used to solve the regression of the interface shape with the different interface models. Let $h(r;\mathbf{w}(t))$ (see Figure \ref{fig:sketch}) denotes the interface profiles solving problems \eqref{eq:quasi-static}, \eqref{eq:quasi-steady} or \eqref{eq:dynamic} at time $t$, parametrized by the set of parameters $\mathbf{w}(t)$. We have $\mathbf{w}(t):=\theta(t)$ for the gravity based model \eqref{eq:quasi-static} and the viscous gravity model \eqref{eq:quasi-steady} and $\mathbf{w}(t):=\theta(t), c_t(t), l_i(t)$ for the inertia-corrected model \eqref{eq:dynamic}.

The regression aims at finding the set of parameters $\mathbf{w}(t)$ which allow each model $h(r;\mathbf{w}(t))$ to best approximate the experimental points $(r_i,h_i)$ sampling the interface shape, with $i\in[1,n_p]$ and $n_p$ the number of points. This allows for indirectly obtaining the contact angle at each time step.

To evaluate the accuracy of the interface regression, we use the average of the squared Fréchèt distances between the set of points and the regressed curves. For each of the points $(r_i,h_i)$, the Fréchèt distance to the interface $h(r;\mathbf{w})$ is denoted as $(d_{\epsilon}(x_i,y_i,h(r_i,\mathbf{w}))$. This is the radius of the smallest circle, centered in $(r_i,h_i)$ and tangent to the curve $h(r;\mathbf{w})$. The use of the Fréchèt distance allows for giving equal importance to errors and uncertainties along both axes, and this is essential in regions near the wall, where $\theta\rightarrow 0$ and thus $|h_r|\rightarrow \infty$. More generally, the Fréchèt distance is particularly useful for regressing functions where the modulus of the derivative is large (see e.g. \cite{Mendez2021_F}).

The best set of weights, for each model, is thus the one that minimizes the objective function

\begin{equation}
\label{eq:cost}
  \underset{\mathbf{w}}{\mathrm{argmin}}\, J(\mathbf{w})=\frac{1}{n_p}\sum_{i=1}^{n_p} \Bigl( d_{\epsilon}(r_i,h_i,h(r_i;\mathbf{w})\Bigr)^2\,.
\end{equation}

\begin{figure}[h!]
    \centering
    \includegraphics[width=9cm, trim={0cm 0cm 0cm 1.5cm},  clip]{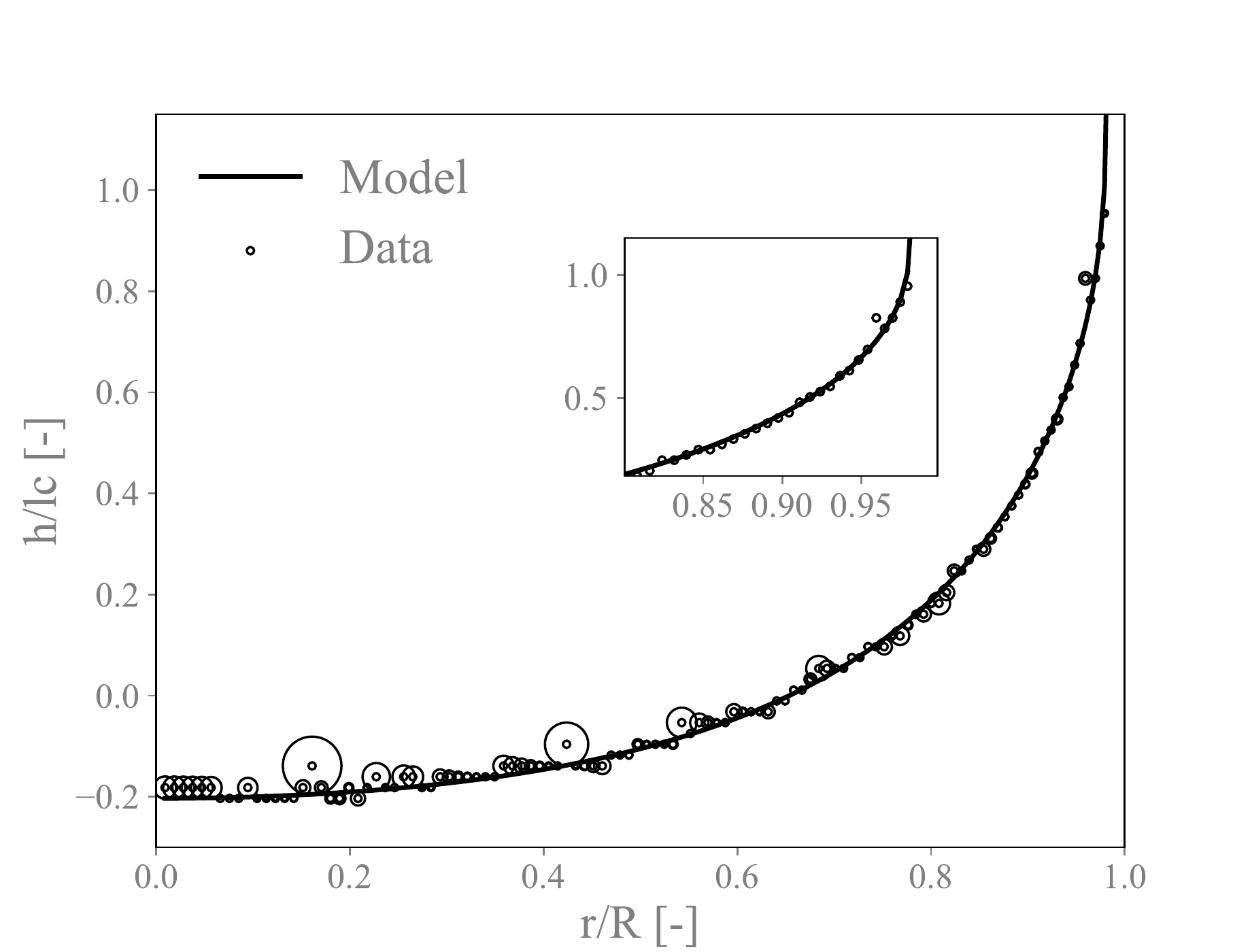}
    \caption{Visualization of the regression based on Fréchèt distance in experimental image. The radius of each circle corresponds to the Fréchèt distance $d_{\epsilon,j}$ of the data point (filled circle) from the model (solid line). The regression aims at finding the model parameters that minimize the function $J(\mathbf{w})$ in \eqref{eq:cost}.}
    \label{fig:Frechet}
\end{figure}

Figure \ref{fig:Frechet} shows an example of interface regression with the circles in each point illustrating the local Fréchèt distance. The inset allows for better appreciating the accuracy of the fit near the wall.

The optimization was carried out using the Nelder-Mead algorithm \cite{nelder1965simplex}, implemented in Python library scipy.optimize \cite{2020SciPy-NMeth}. For each model, we used the boundary value problem solver in scipy.integrate \cite{2020SciPy-NMeth}; this implements a collocation algorithm coupled with a damped Newton method \cite{ascher1995numerical}. The optimization space for the weights was identified by trial and error and the optimization is carried out on a normalized domain, i.e. keeping all weights in the range $[0,1]$ using a min-max scaling.

Uncertainties are computed using a bootstrapping approach \cite{efron1994introduction}. For each experimental image, the regression is repeated $n_b=500$ times using $70\%$ of the available $n_p$ points. Each time, the points used in the regression are sampled randomly. 
The results of the multiple regressions produce a distribution of weights $\mathbf{w}\sim p(\mathbf{w})$. Since these were approximately Gaussian, the uncertainty in each parameter is computed as 1.96 times the (sample) standard deviation of the weight distribution, thus considering a 95\% confidence interval. The uncertainty in the interface detection is then computed using a Monte Carlo approach: the distributions of weights are propagated by solving the model equations $n_b$ times and computing the associated uncertainty bounds from the standard deviations of the corresponding interface $h(r;\mathbf{w})$ distributions.

\subsection{Validation of the image analysis method}\label{subsec:validation}

We analyzed the accuracy of the proposed MPM method by testing it on a set of meniscus profiles for which the ground-truth is available. To this end, we use the IC model \eqref{eq:dynamic} to generate a set of profiles by varying the model parameters $\mathbf{w}:=c_t,l_i,\theta$ and $Ca$ within the ranges observed in the experiments. This means $c_t\in[-1,1]$, $l_i\in[4.e-4,4.e-3]$ m, $\theta\in [5^\circ,45^\circ]$ and $Ca\in [-5.e-3,5.e-3]$. We randomly sampled 30 points using a uniform distribution within these ranges and generate the associated interface profiles $h(r;\mathbf{w},Ca)$ by solving \eqref{eq:dynamic} on a domain $r\in[0,(R-l_m )]$, where $l_m$ is the smallest distance from the wall at which it is possible to sample the interface location. Noticing that in our experiments $l_m$ varies in the range $[1,5]$ $\mu$m, we consider $l_m=(1,10,30)$ $\mu$m, to analyze the impact of the image resolution in the interface regression. Moreover, we add random noise to the interface location to simulate an experiment. This noise was uniformly distributed, with zero mean and $10 \mu$m standard deviation. The parameter uncertainty was computed using the bootstrapping approach discussed in section \ref{subsec:inverse_method}. We compare this approach with the results obtained with standard tangent line method (TLM). In this case, for each of the 30 generated profiles, we perform a linear extrapolation of the interface using randomly sampled interface points among the last 100 $\mu m$ to the wall. For  each measurements we generate a Gaussian distribution of contact angles by repeating the extrapolation $n_{tlm}=500$ times using 70\% of the available points. The contact angle is obtained from the mean of the contact angle distribution, with the uncertainty computed as 1.96 times the standard deviation.

\begin{figure}
    \centering
    \includegraphics[width=8.5cm, clip]{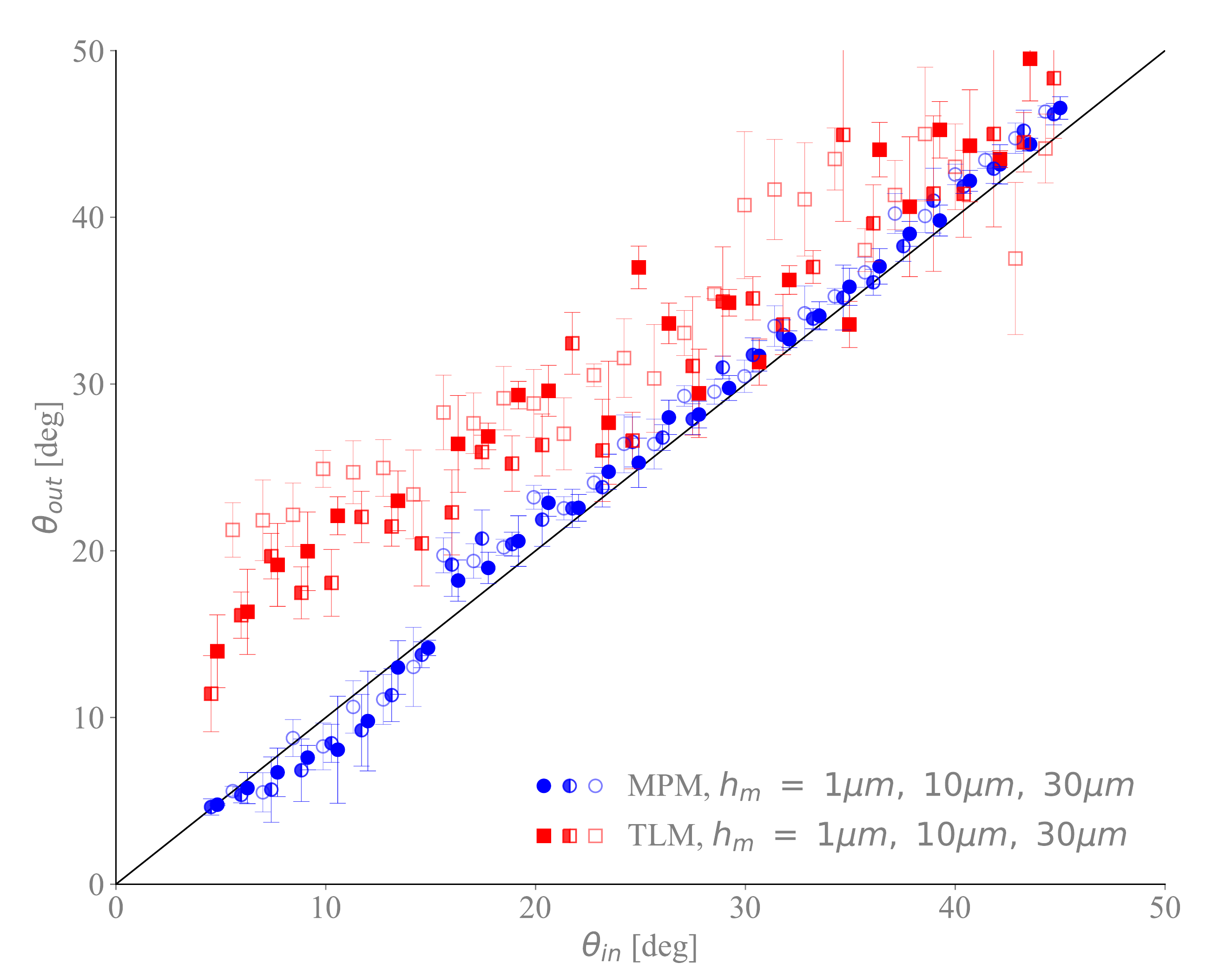}
    \caption{Accuracy comparison between two different methods, MPM (circles) and TLM (squares), for contact angle quantification on synthetic images by using different degree of vicinity to the contact line: $l_m= 1\mu m$ (full symbols), $l_m= 10\mu m$ (left-filled symbols) and $l_m= 30\mu m$ (empty symbols).}
    \label{fig:SyntheticAnalysis}
\end{figure}

The results of this analysis are shown in Figure \ref{fig:SyntheticAnalysis}. This plots the contact angle derived via the MPM and TLM approach as a function of the actual contact angle in the generated interfaces. The results obtained using the MPM are plotted with circles, with different markers filling distinguishing profiles with the three levels of interface resolution at the wall. The Figure also reports, with red squares, the results obtained using the TLM.

The result shows that the TLM significantly over-predicts the contact angle when this is below $15^\circ$; this result is in line with those reported by \citet{maleki2007landau}. The interface resolution's impact becomes more pronounced at $\theta <15^\circ$, while the proposed MPM is more accurate and more robust against poor interface resolution over the entire range of parameters and analyzed operating conditions. Nevertheless, it is worth noticing that the uncertainties in the contact angle determination grow with smaller contact angles and reach up to $5^\circ$ in some conditions, regardless of the interface resolution.

\section{Results}\label{sec:results}

\begin{figure*}
    \centering
    \begin{subfigure}[t]{0.47\textwidth}
    \includegraphics[width=8.5cm, trim={0cm 0.cm 0cm 0.cm},  clip]{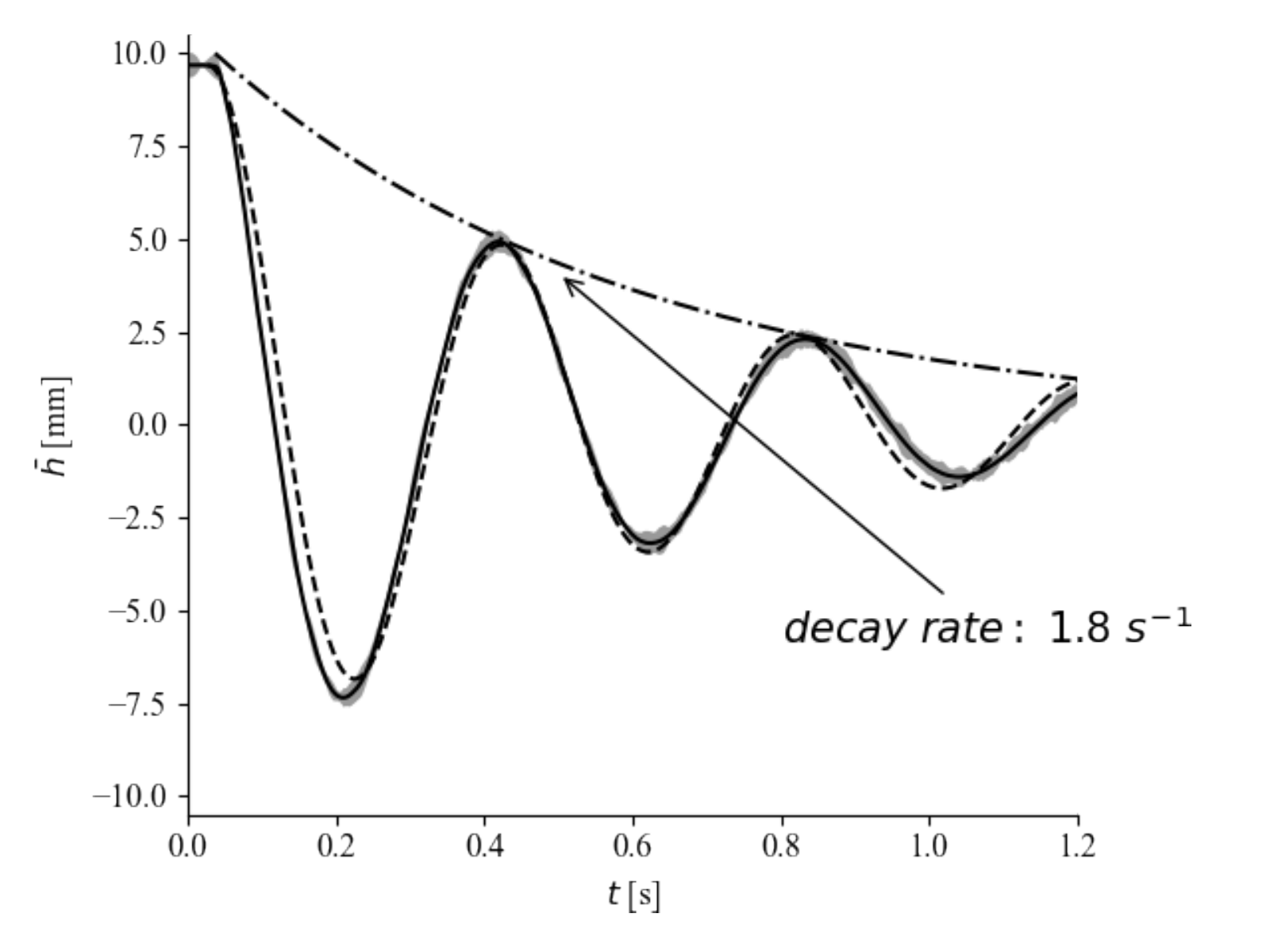}
    \caption{Interface displacement for HFE7200}
    \label{fig:HFEexperiment}
    \end{subfigure}
    \hfill
    \begin{subfigure}[t]{0.47\textwidth}
    \includegraphics[width=8.5cm, trim={0cm 0.cm 0cm 0.cm},  clip]{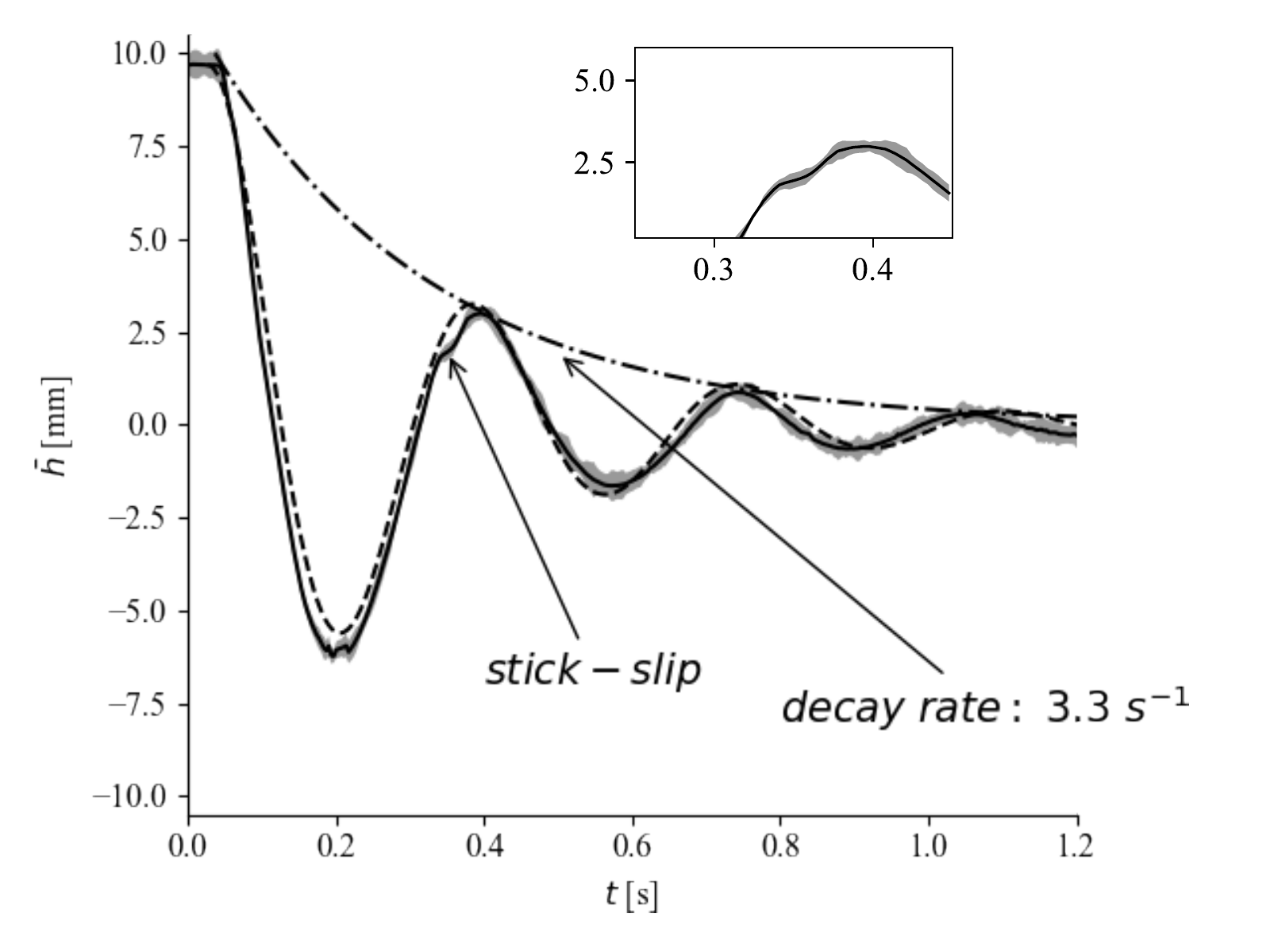}
    \caption{Interface displacement for demineralized water}
    \label{fig:WATERexperiment}
    \end{subfigure}
    \hfill
    \begin{subfigure}[t]{0.47\textwidth}
    \includegraphics[width=8.5cm, trim={0cm 0cm 0cm 0cm},  clip]{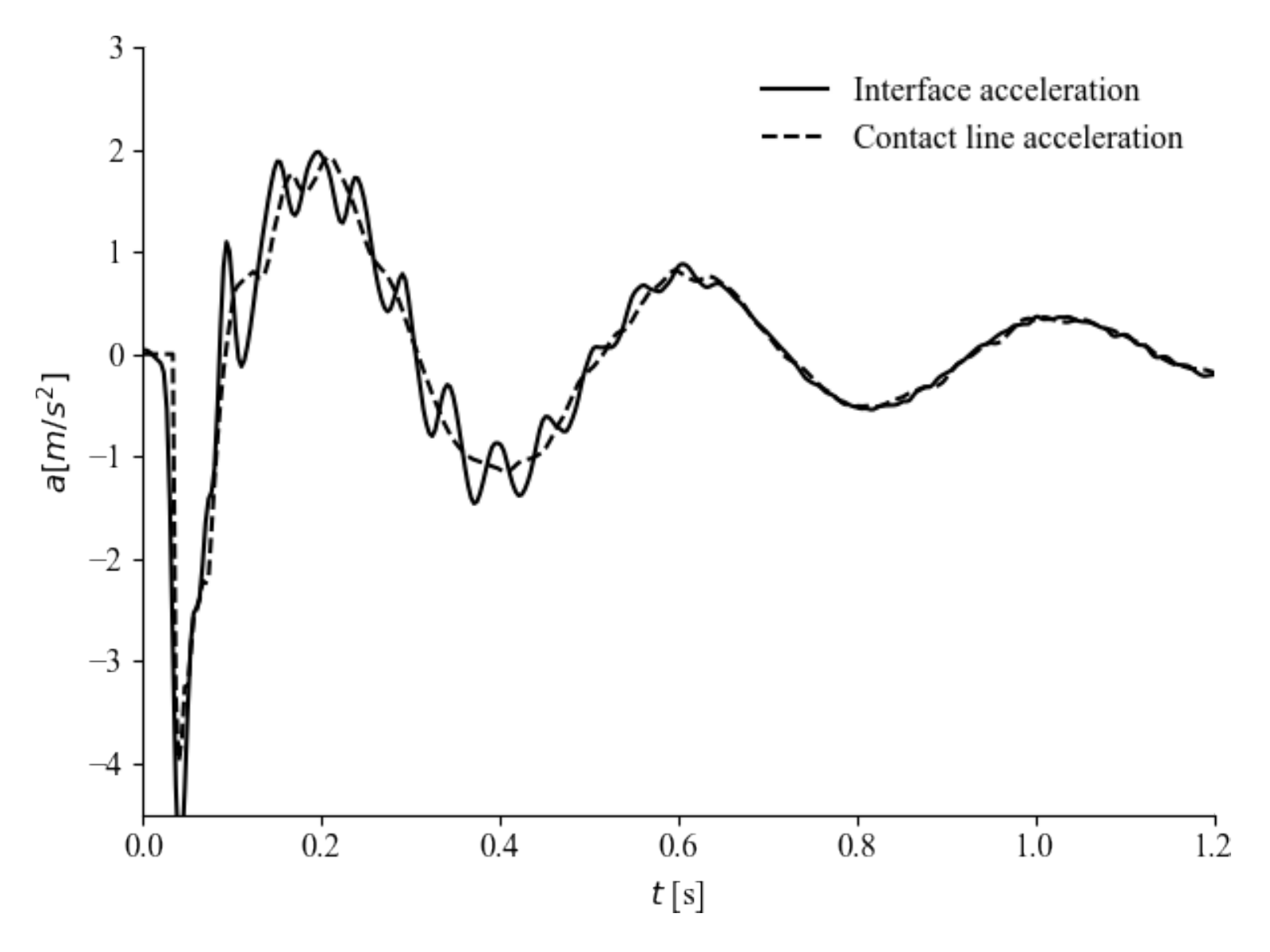}
    \caption{Interface accelerations for HFE7200}
    \label{fig:HFEacceleration}
    \end{subfigure}
    \hfill
    \begin{subfigure}[t]{0.47\textwidth}
    \includegraphics[width=8.5cm, trim={0cm 0cm 0cm 0cm},  clip]{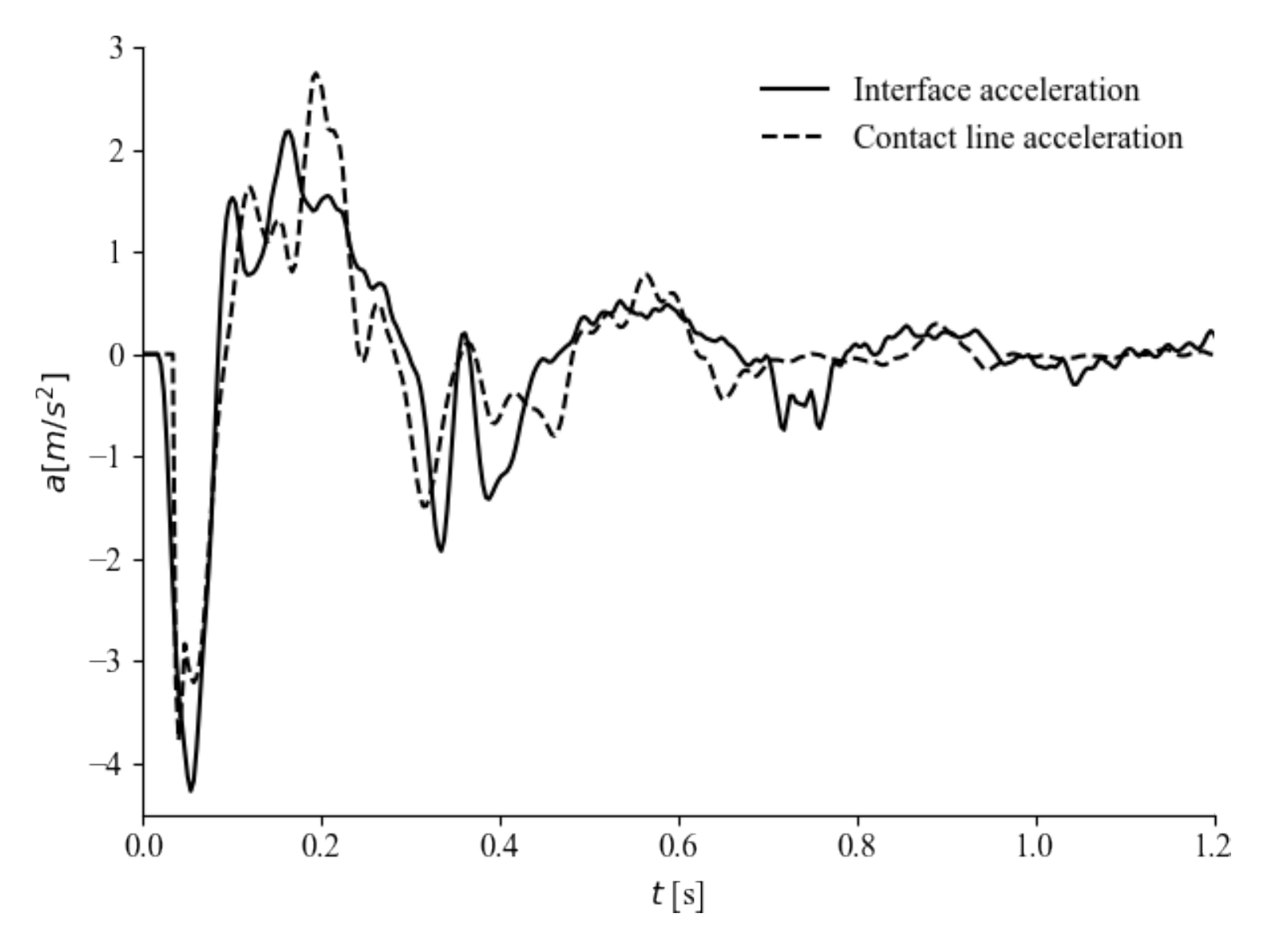}
    \caption{Interface accelerations for demineralized water}
    \label{fig:WATERacceleration}
    \end{subfigure}
    \caption{Figures a-b show the evolution of the mean interface height $\overline{h}$ for a test case in HFE7200 and demineralized water. The dashed line shows the response of a canonical second order system with $\omega_n = 15.96$ rad/s, $\zeta = 0.11$ for HFE7200 and $\omega_n=17.8$ rad/s and $\zeta=0.17$ for water, where the zoom view highlights the stick-slip motion of the contact line. The dashed-dot line shows the exponential envelope for the oscillation decay ($e^{-\lambda t}$).
    Figures c-d  show the acceleration of the spatial averaged interface (solid line) and contact line (dashed). In the case of water (\ref{fig:WATERacceleration}) the behaviour is chaotic because of the stick slip of the contact line.}
\end{figure*}

We begin the analysis by illustrating the evolution of the liquid height as a function of time on one side of the tube for both liquids. Two representative examples are shown in Figure \ref{fig:HFEexperiment} for HFE7200 and Figure \ref{fig:WATERexperiment} for demineralized water. 

\begin{figure}[h]
    \centering
    \begin{subfigure}[t]{0.2\linewidth}
    \includegraphics[width=1\textwidth]{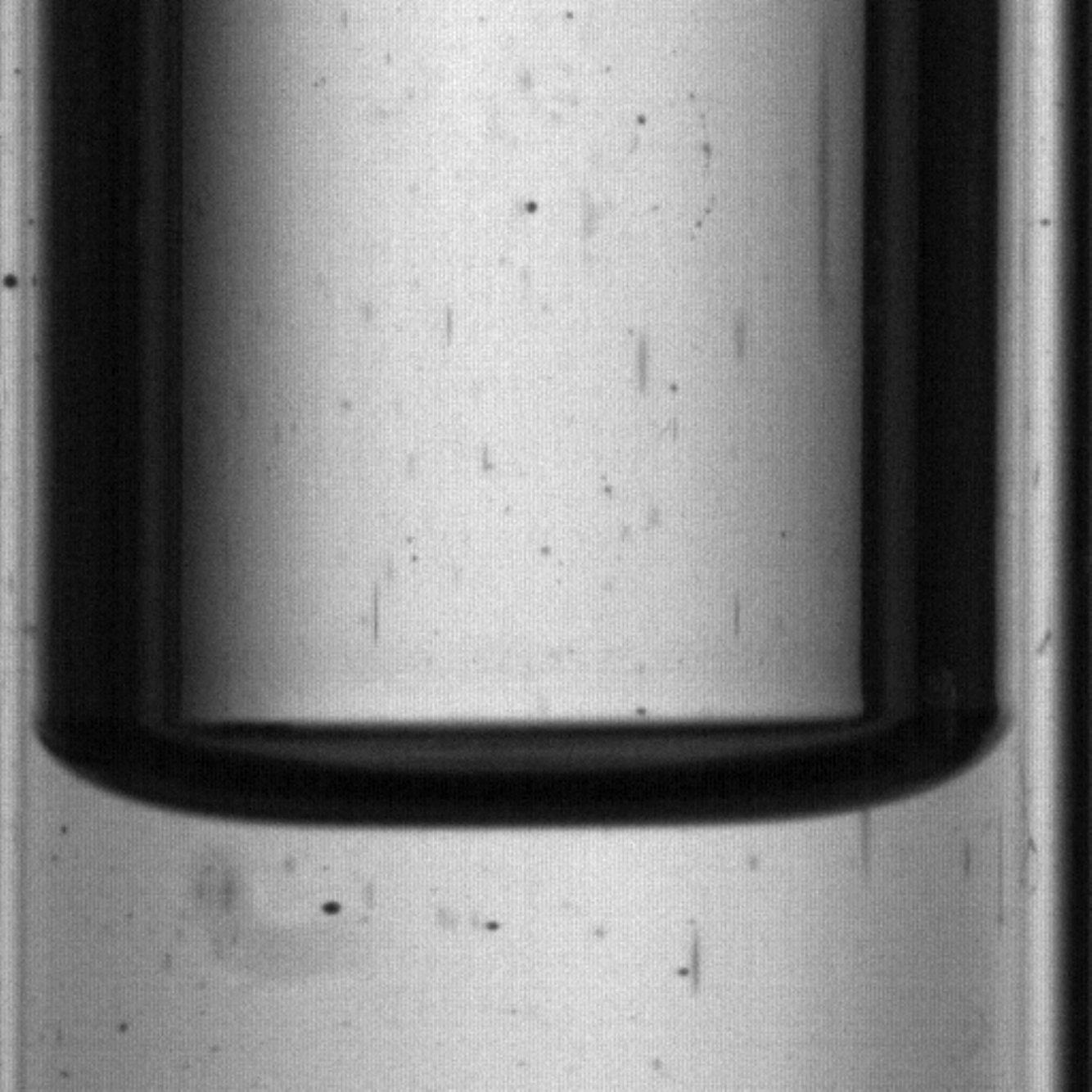}
    \caption{$0.270s$}\label{fig:hfeSnapshotsA}
    \end{subfigure}%
    \hfill
    \begin{subfigure}[t]{0.2\linewidth}
    \includegraphics[width=1\textwidth]{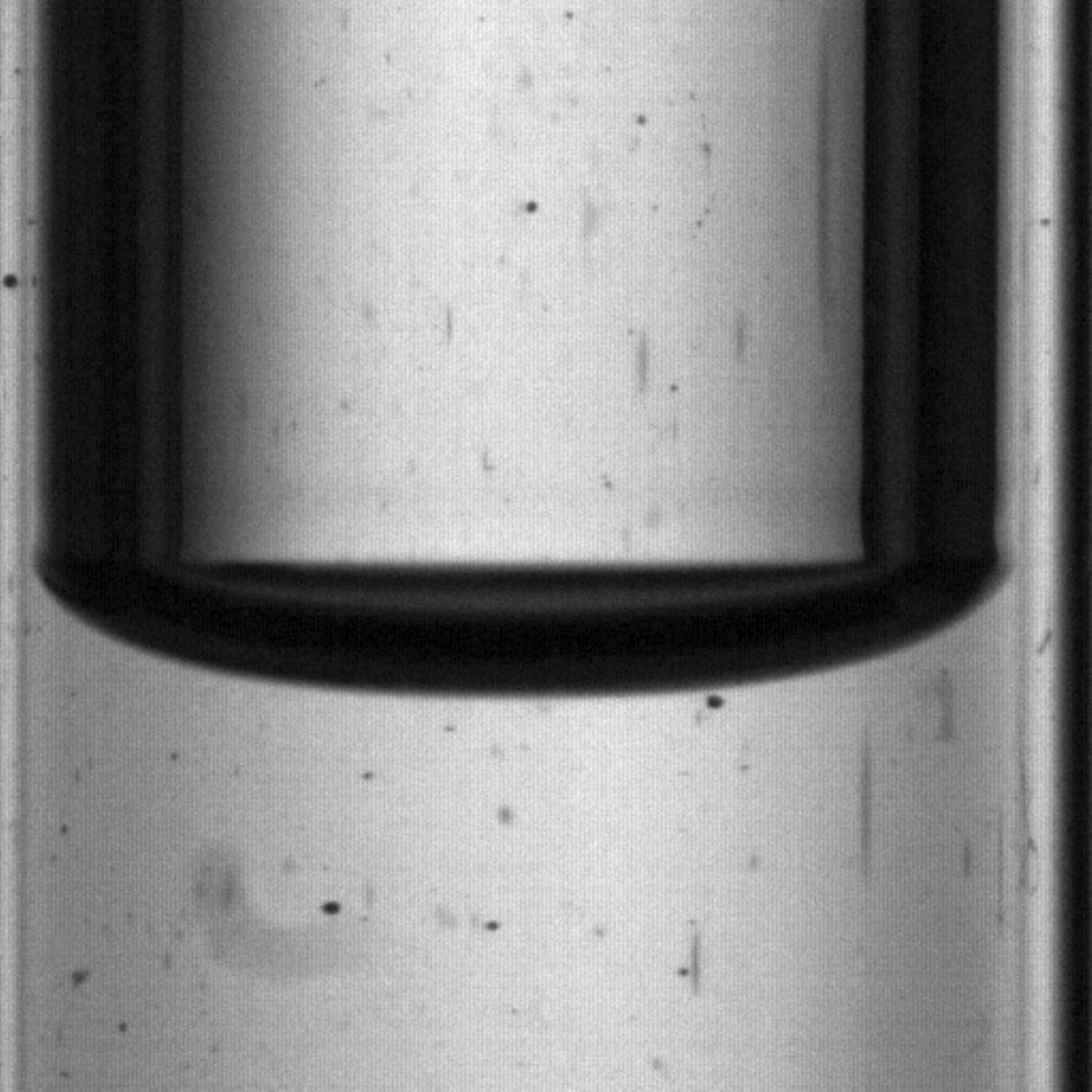}
    \caption{$0.287s$}\label{fig:hfeSnapshotsB}
    \end{subfigure}%
    \hfill
    \begin{subfigure}[t]{0.2\linewidth}
    \includegraphics[width=1\textwidth]{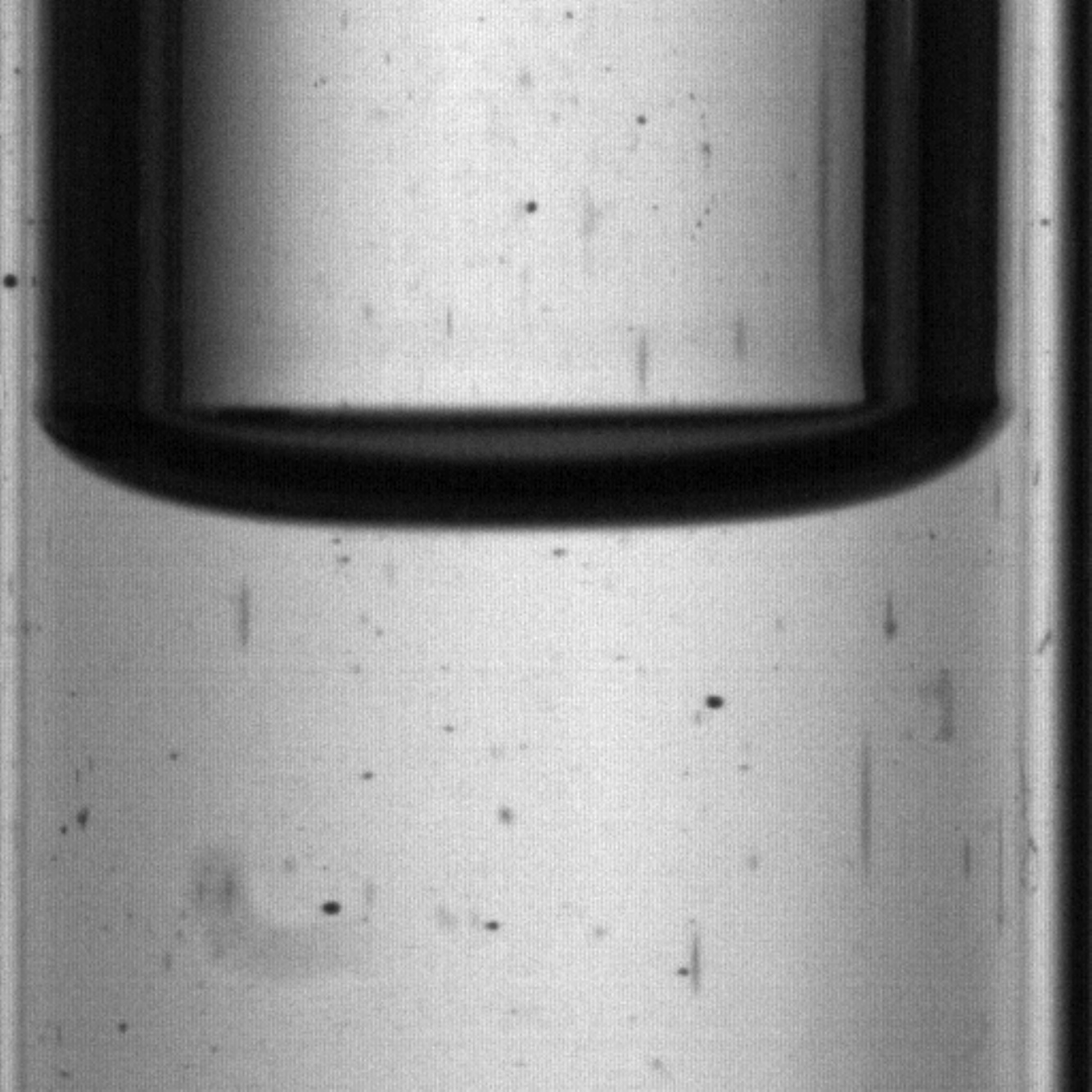}
    \caption{$0.304s$}\label{fig:hfeSnapshotsC}
    \end{subfigure}%
    \hfill
    \begin{subfigure}[t]{0.2\linewidth}
    \includegraphics[width=1\textwidth]{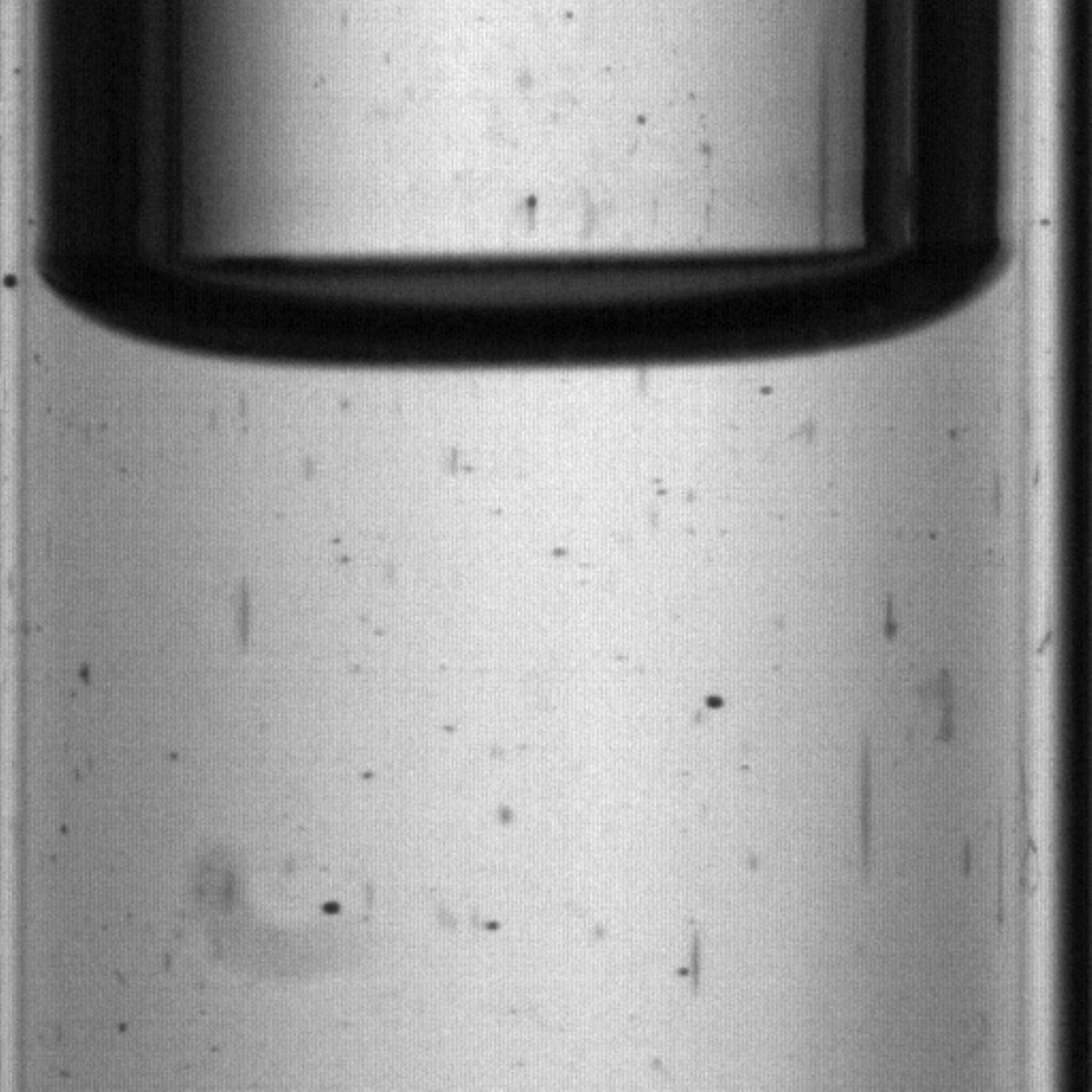}
    \caption{$0.320s$}\label{fig:hfeSnapshotsD}
    \end{subfigure}%
    \hfill
    \begin{subfigure}[t]{0.2\linewidth}
    \includegraphics[width=1\textwidth]{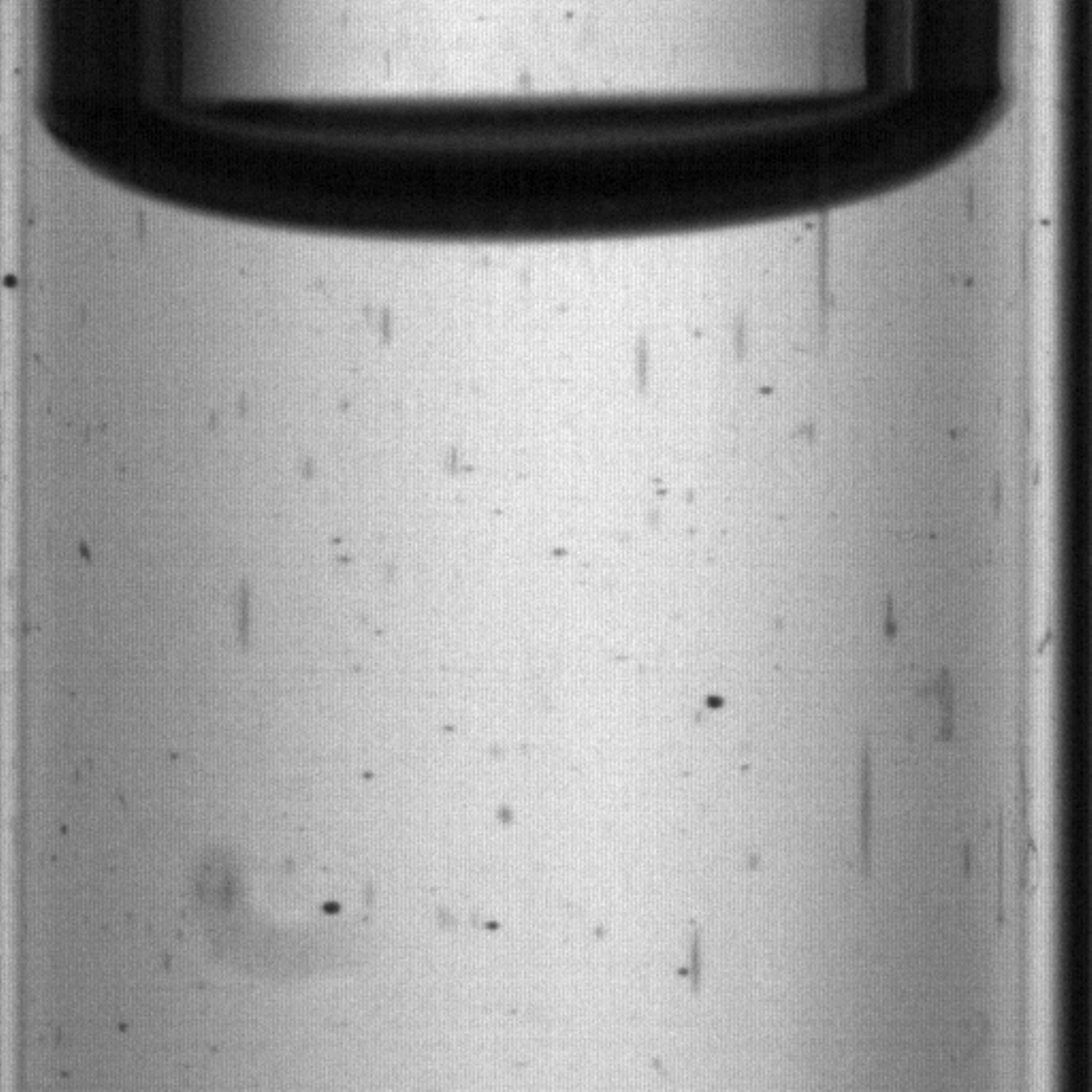}
    \caption{$0.337s$}\label{fig:hfeSnapshotsE}
    \end{subfigure}%
    \hfill
    \begin{subfigure}[t]{0.2\linewidth}
    \includegraphics[width=1\textwidth]{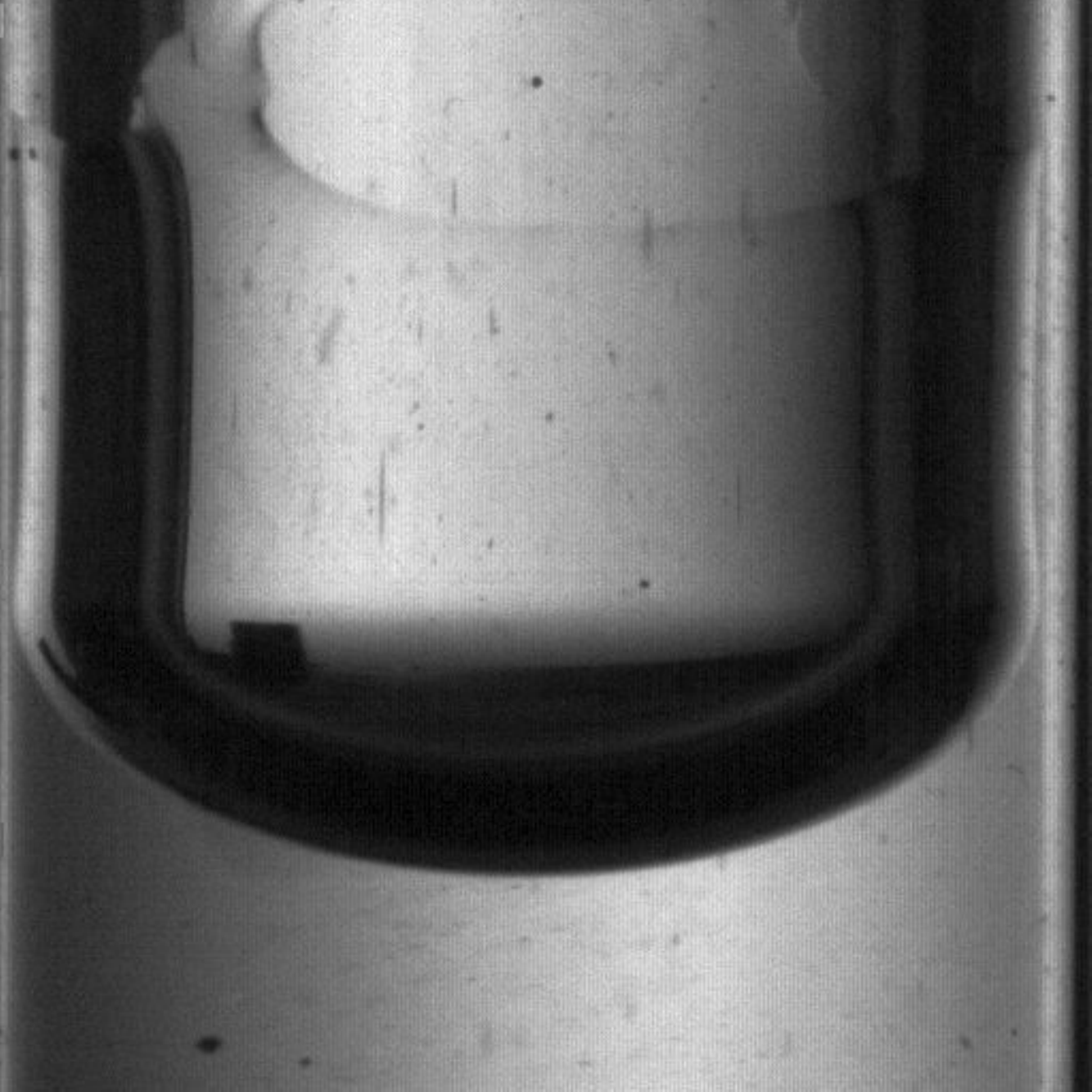}
    \caption{$0.270s$}\label{fig:waterSnapshotsA}
    \end{subfigure}%
    \hfill
    \begin{subfigure}[t]{0.2\linewidth}
    \includegraphics[width=1\textwidth]{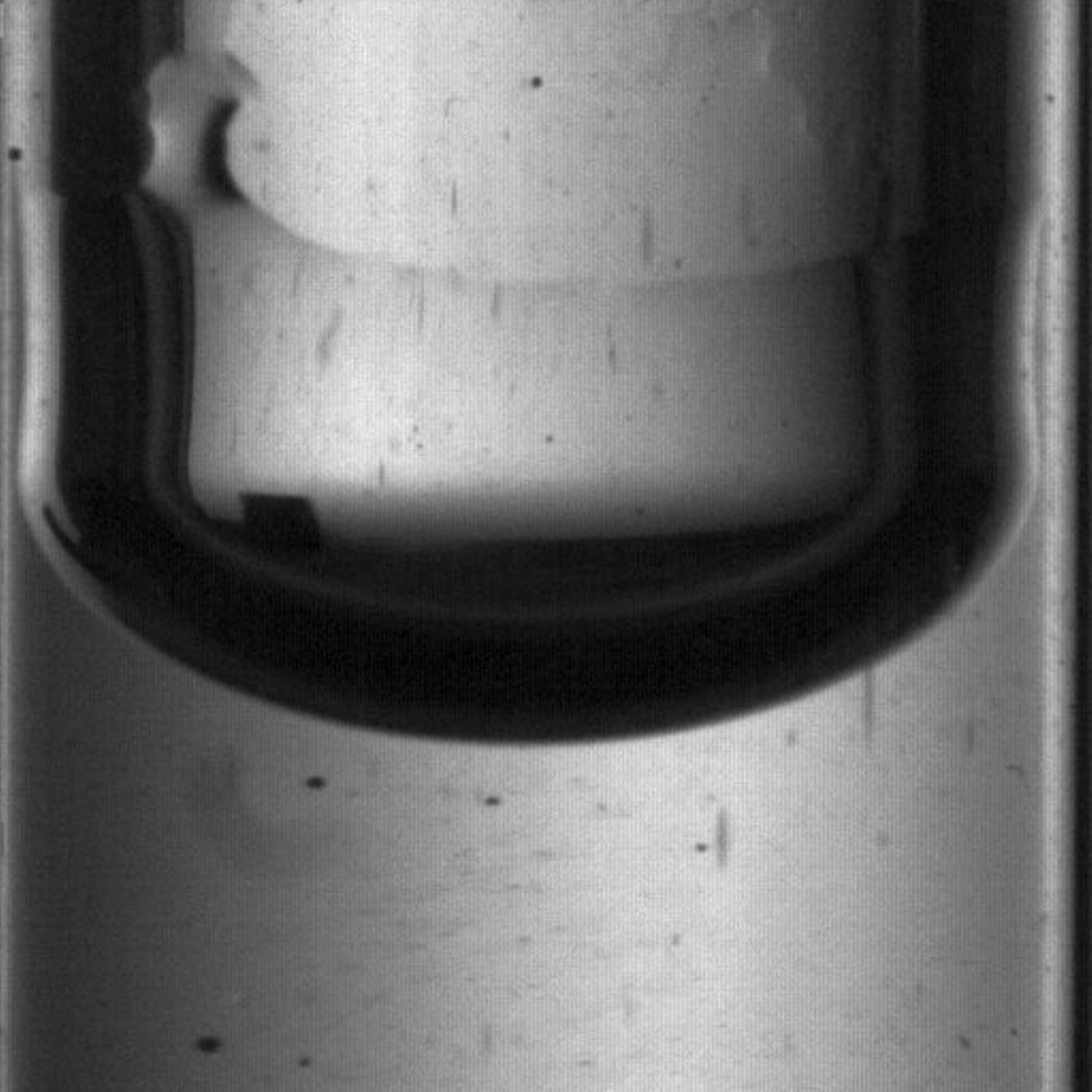}
    \caption{$0.287s$}\label{fig:waterSnapshotsB}
    \end{subfigure}%
    \hfill
    \begin{subfigure}[t]{0.2\linewidth}
    \includegraphics[width=1\textwidth]{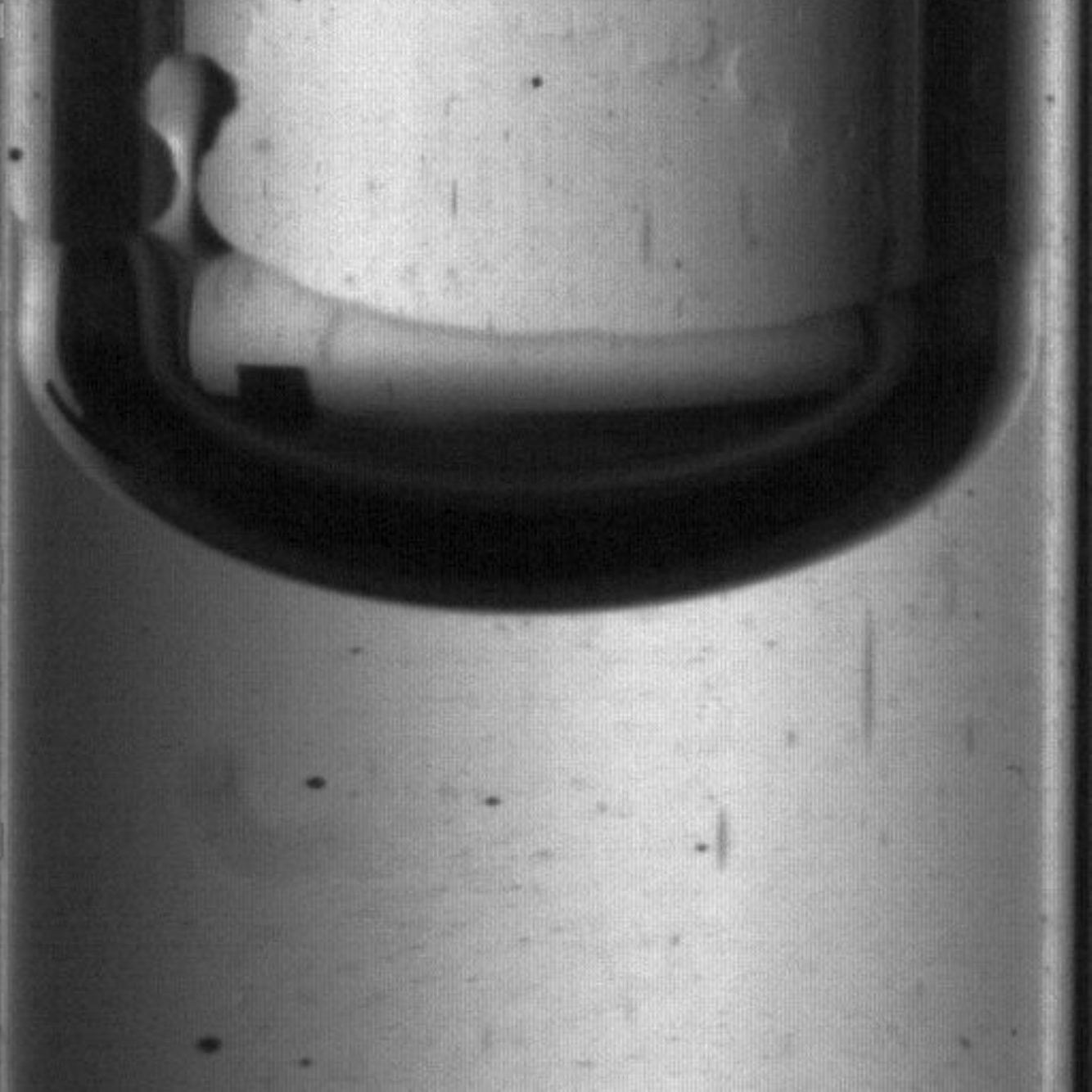}
    \caption{$0.304s$}\label{fig:waterSnapshotsC}
    \end{subfigure}%
    \hfill
    \begin{subfigure}[t]{0.2\linewidth}
    \includegraphics[width=1\textwidth]{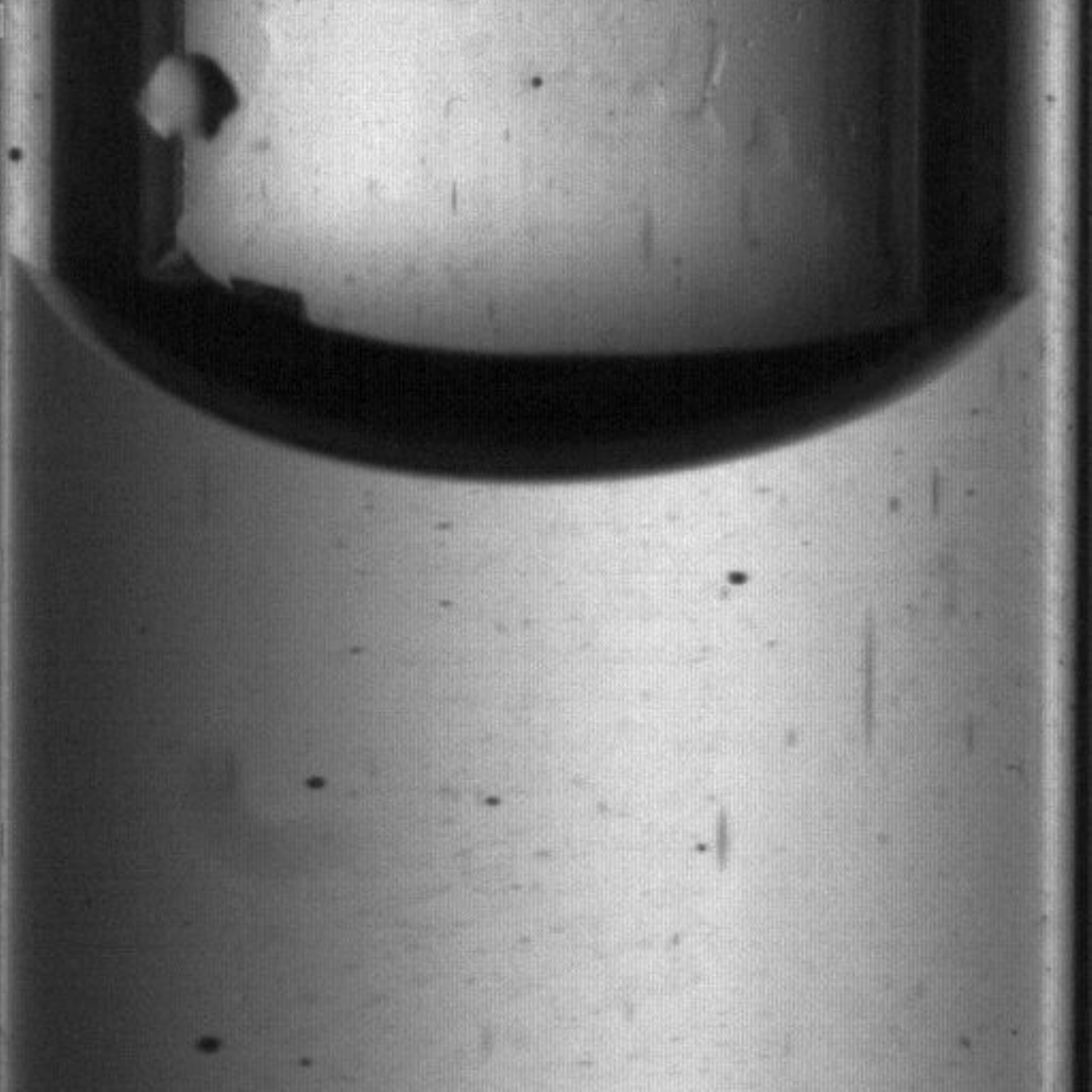}
    \caption{$0.320s$}\label{fig:waterSnapshotsD}
    \end{subfigure}%
    \hfill
    \begin{subfigure}[t]{0.2\linewidth}
    \includegraphics[width=1\textwidth]{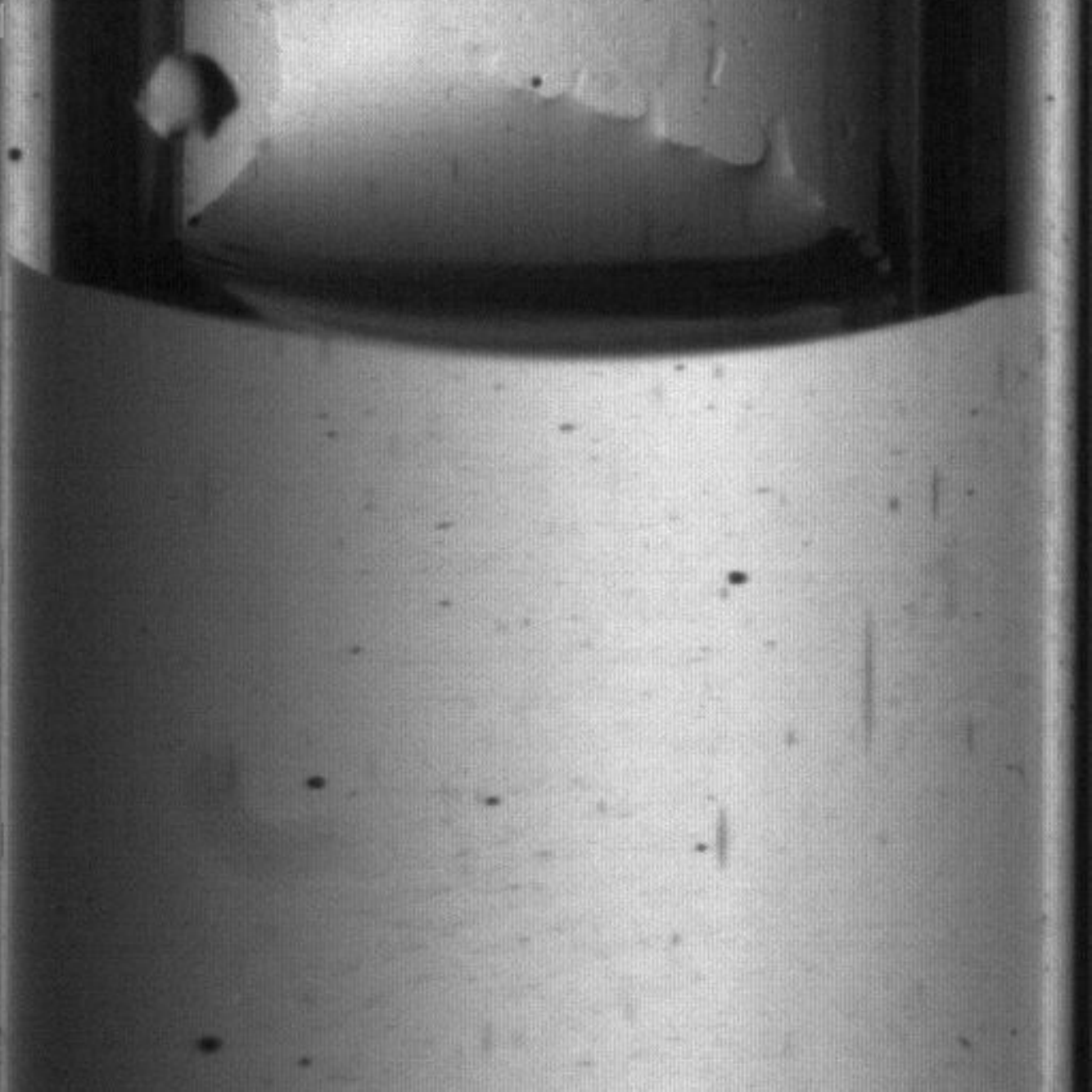}
    \caption{$0.337s$}\label{fig:waterSnapshotsE}
    \end{subfigure}%
    \caption{Snapshots of the experiment with HFE7200 (a-b-c-d-e) and demineralized water (f-g-h-i-j) in 8 mm diameter tube. In \ref{fig:waterSnapshotsD} the rising interface encounters the descending film. The slope of the interface near the wall suddenly increases and the interface stops its motion.}
    \label{fig:waterSnapshots}
\end{figure}

The height $\bar{h}$ in these plots is the spatial average over the detected menisci, shifted so that $h=0$ corresponds to the equilibrium position. Each experiment starts with the same height of $\bar{h}=10$ mm and the interface position oscillates within a range of $\approx 17.5$ mm of axial distance within the tube. The length of the fluid column controls the (natural) frequency of the oscillation, which for the experiments presented here is $2.5 $ Hz for HFE7200 (Figure \ref{fig:HFEexperiment}) and $2.7$ Hz for demineralized water (Figure \ref{fig:WATERexperiment}). 

The gray area indicates the uncertainty on the average height computed from the interface detection (see supplementary material). These experiments were repeated three times for repeatability analysis; since the standard deviation between various experiments is generally lower than the uncertainties in the interface detection within a single run, we hereby consider only one representative test case for each fluid. For both fluids, the response of the liquid interface qualitatively behaves as a second-order canonical linear dynamical system of the form $\ddot{\bar{h}}+2\zeta \omega_n \dot{\bar{h}}+\omega^2_n \bar{h}$, which one could derive from a simple macroscopic model for the liquid column dynamics. 

The decay rate computed from the experimental data is also indicated in Figures \ref{fig:HFEexperiment} and \ref{fig:WATERexperiment}, and its exponential envelope is shown with dash-dot lines. The dashed line illustrates the prediction of the `best' canonical second-order system for the data at hand, identified via regression of the natural pulsation $\omega_n$ and damping factor $\zeta$. The results of the regression are indicated for each fluid in the captions of Figures \ref{fig:HFEexperiment} and \ref{fig:WATERexperiment}.
The matching is overall satisfactory if one considers the simplification involved in modelling the interface dynamics using an autonomous dynamical system.

The higher viscosity and higher capillary forces in the case of water lead to 1.8 times higher decay rate. Moreover, in the case of water, the contact line systematically pins to the wall a few instants before reaching the second peak. The snapshots of Figure \ref{fig:waterSnapshots} show that this condition occurs when the rising interface meets the liquid film deposited at the previous descend.  Figures \ref{fig:waterSnapshotsA}-\ref{fig:waterSnapshotsC} show that the film moves towards the interface prior to the pinning, which is shown by Figure \ref{fig:waterSnapshotsD}. After the pinning the velocity of the contact line is null, but the liquid column pushes upward the central part of the interface as shown in Figure \ref{fig:waterSnapshotsE}. The associated `stick-slip' motion of the contact line slows down the rising of the liquid interface (see annotation in Figure \ref{fig:WATERexperiment} and the zoomed view around $t\in [0.3-0.4]$s). This dynamic is not observed in the case of HFE7200. 

The interface deformation and oscillations in these experiments were observed to be fairly axial-symmetric. In the case of water, the film deposition on the wall alters the axisymmetry of the interface (cf. figures \ref{fig:waterSnapshotsA}-\ref{fig:waterSnapshotsB}-\ref{fig:waterSnapshotsC}) but this condition lasts only few milliseconds and does not appreciably impact the interface as shown by the figures \ref{fig:waterSnapshots}. 

Two main mechanisms could be qualitatively identified. The first mechanism arguably originates at the wall: as the contact line velocity changes, the contact angle changes and with it, at least up to some distance from the wall, the interface shape. The second mechanism originates at the channel's center: because the flow far from the wall moves faster, its acceleration involves larger inertial contributions. These triggers interface oscillations, as shown in Figures \ref{fig:HFEacceleration} and \ref{fig:WATERacceleration} for HFE7200 and water respectively. These Figures show the acceleration of the spatially averaged interface height $\ddot{\bar{h}}(t)$ and the acceleration of the contact line ($\ddot{h}(R,t)$). In both cases, the time differentiation is carried out after smoothing the time series using a Savitzky–Golay filter.

In the case of HFE7200 (Figure \ref{fig:HFEacceleration}), the interface oscillates significantly at the center of the channel, at a frequency ($\approx 20$Hz) much larger than the frequency of the liquid column's oscillation. This oscillation begins right after the first minima and lasts until $t=0.5$s.
In the case of water (Figure \ref{fig:WATERacceleration}), the acceleration profiles are somewhat chaotic because of the stick-slip motion of the contact line. Nevertheless, these oscillations can be seen within the first $0.4$s.

While in both fluids these oscillations are too small to be visible in the average interface height $\bar{h}(t)$ in Figure \ref{fig:HFEexperiment} and \ref{fig:WATERexperiment}, their impact on the contact angle is significant depending on which model is used for the interface regression, as discussed in the following section.

\subsection{Experimental validation of Interface models}\label{subsec:result_interfaces}

\begin{figure*}
    \centering
    \begin{subfigure}[t]{0.91\textwidth}
    \includegraphics[width=0.99\textwidth, trim={2cm 3.5cm 3cm 4.3cm},  clip]{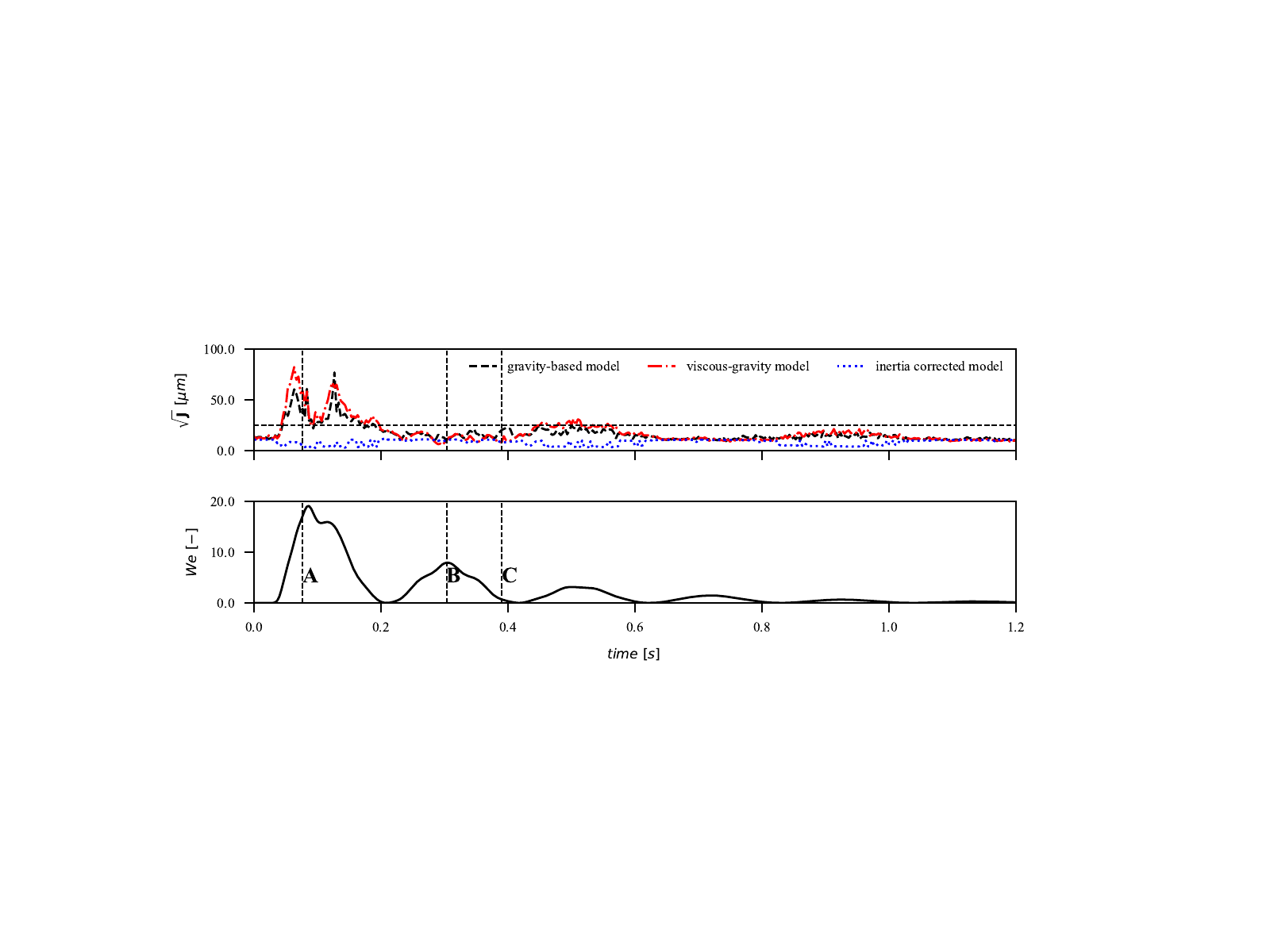}
    \caption{Regression error and Weber number history for the experiment with HFE7200}
    \label{fig:HFE regression error}
    \end{subfigure}
    \begin{subfigure}[t]{0.91\textwidth}
        \includegraphics[width=0.99\textwidth, trim={2cm 3.5cm 3cm 4.3cm},  clip]{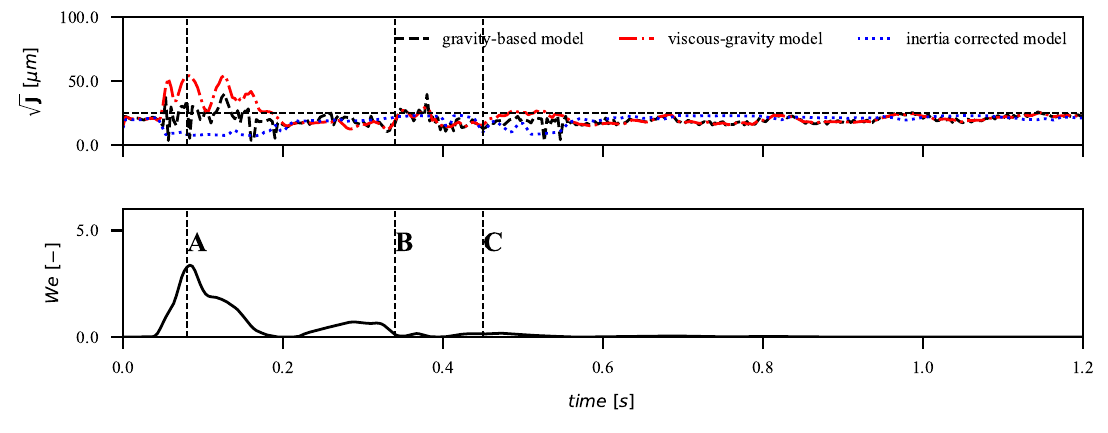}
    \caption{Same as Figure \ref{fig:HFE regression error} but considering a test case with demineralized water}
    \label{fig:WATER regression error}
    \end{subfigure}
    \caption{The Figure on the top shows the regression error as a function of time for the three models. The Figure on the bottom shows the Weber number of the interface as a function of time and the three example conditions (A-B-C) at which the interface is later analyzed in the what follows.}
\end{figure*}

\begin{figure*}
    \centering
    \begin{subfigure}[t]{0.47\textwidth}
    \includegraphics[width=8.5cm, trim={.8cm 1cm 0cm 1cm},  clip]{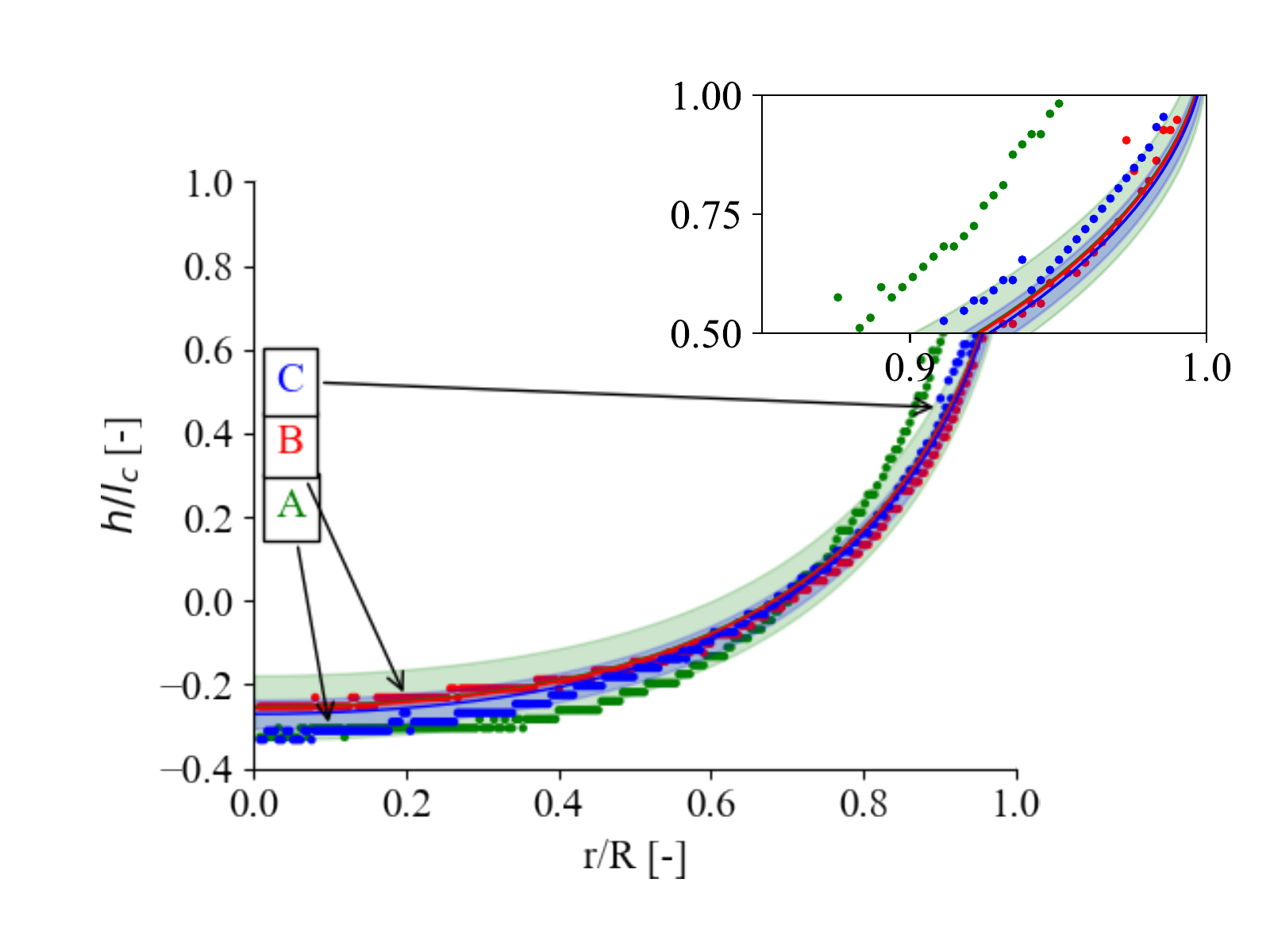}
    \caption{HFE7200 interface vs GB model}
    \label{fig:HFE gravity based}
    \end{subfigure}
    \hfill
    \hspace{-15mm}
    \begin{subfigure}[t]{0.47\textwidth}
    \includegraphics[width=8.5cm, trim={.8cm 1cm 0cm 1cm},   clip]{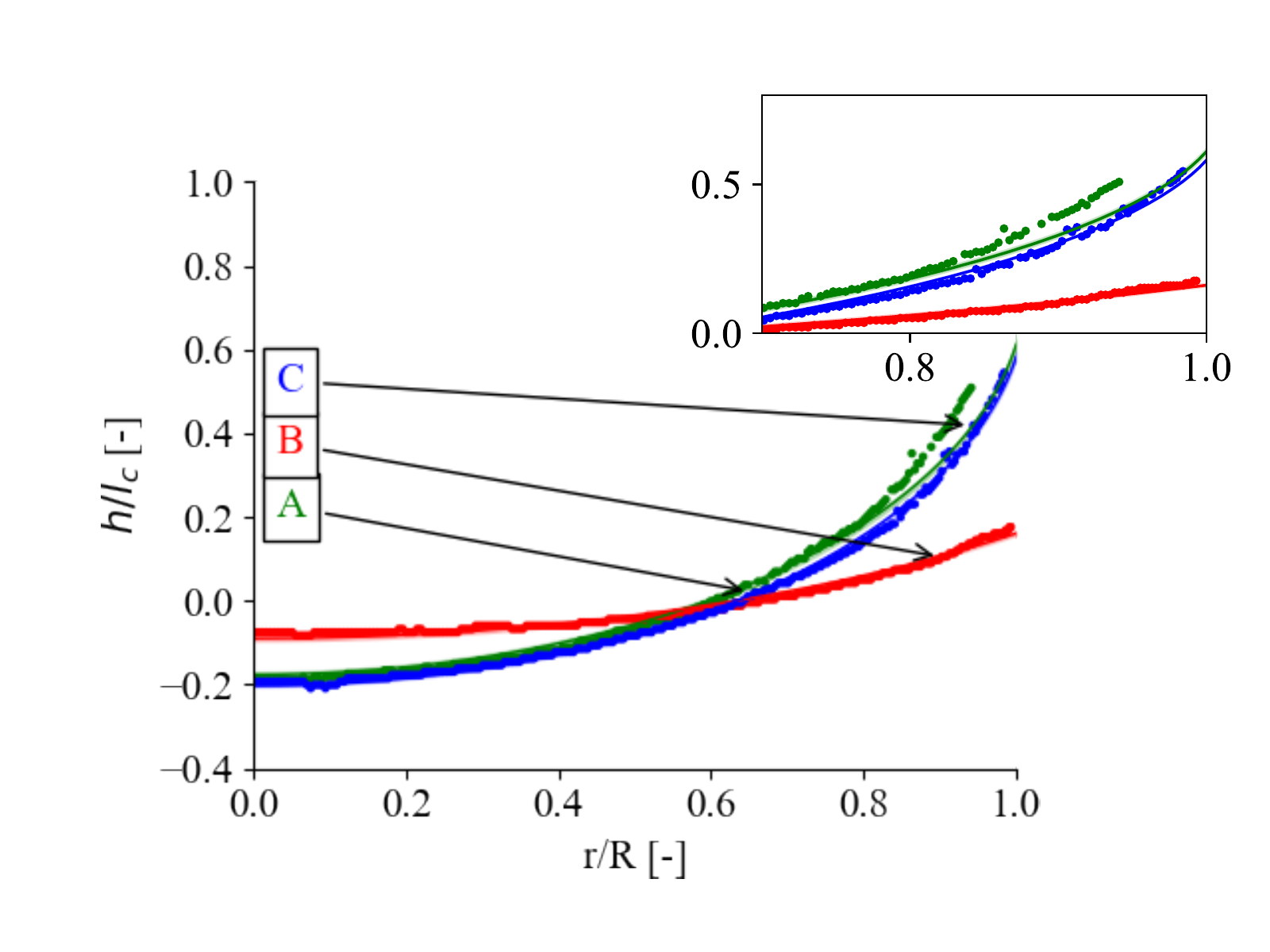}
    \caption{demineralized water interface vs GB model}
    \label{fig:WATER gravity based}
    \end{subfigure}
    \hfill
    \hspace{-15mm}
    \begin{subfigure}[t]{0.47\textwidth}
    \includegraphics[width=8.5cm, trim={.8cm 1cm 0cm 1cm},  clip]{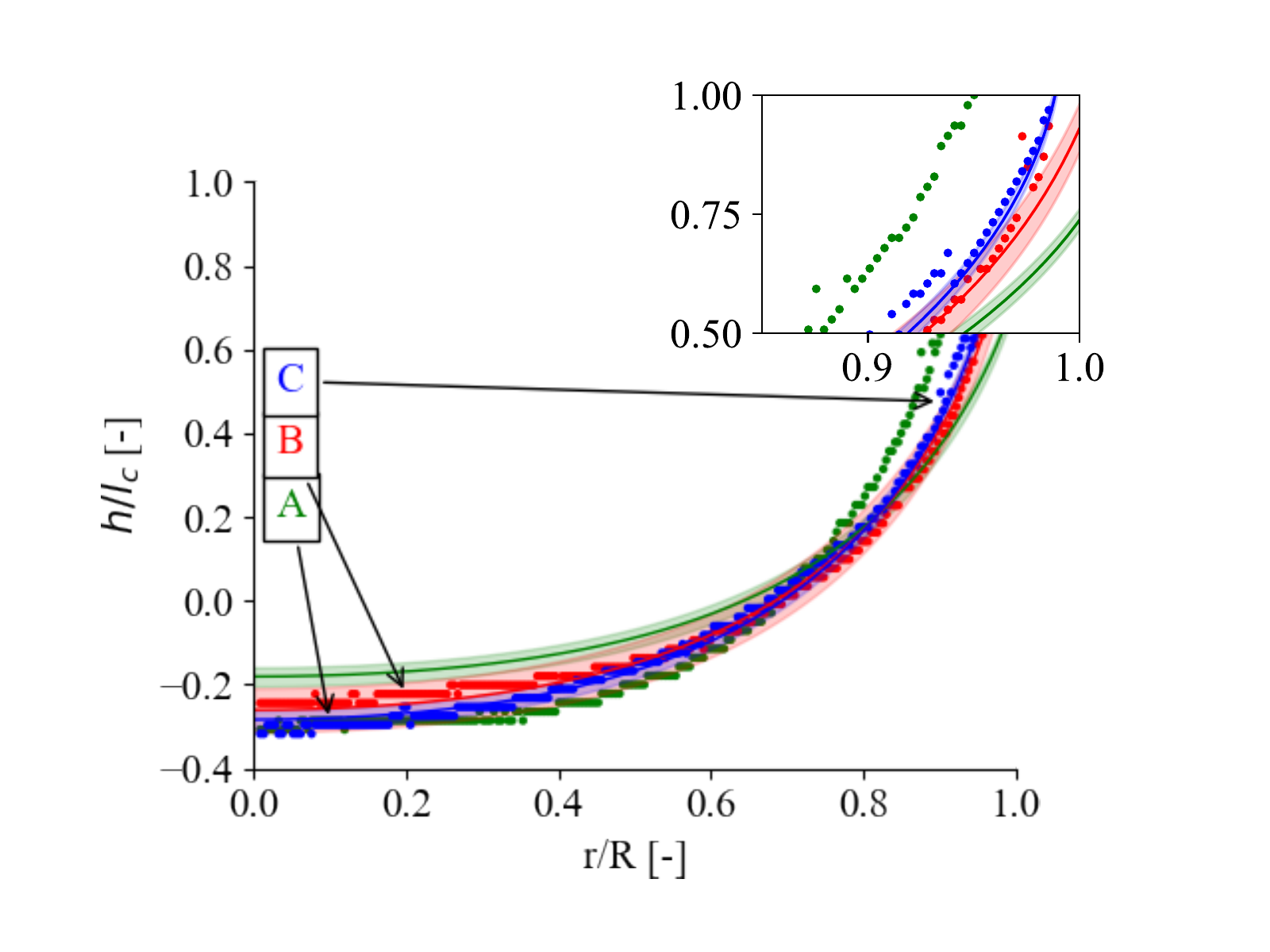}
    \caption{HFE7200 interface vs VG model}
    \label{fig:HFE viscous based}
    \end{subfigure}
    \hfill
    \hspace{-15mm}
    \begin{subfigure}[t]{0.47\textwidth}
    \includegraphics[width=8.5cm, trim={.8cm 1cm 0cm 1cm},  clip]{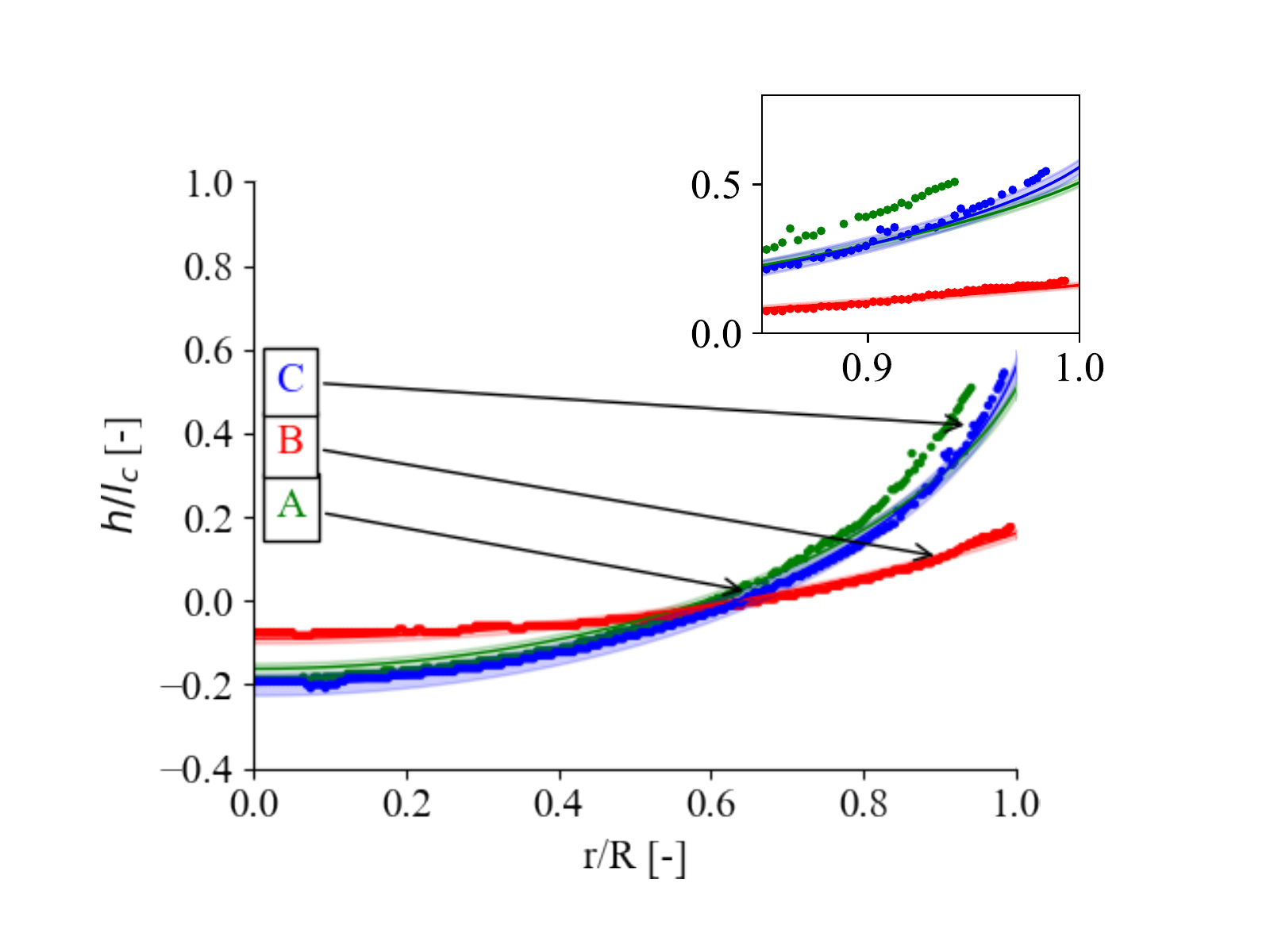}
    \caption{demineralized water interface vs VG model}
    \label{fig:WATER viscous based} 
    \end{subfigure}
    \hfill
    \hspace{-15mm}
    \begin{subfigure}[t]{0.49\textwidth}
    \includegraphics[width=8.5cm, trim={.8cm 1cm 0cm 1cm},   clip]{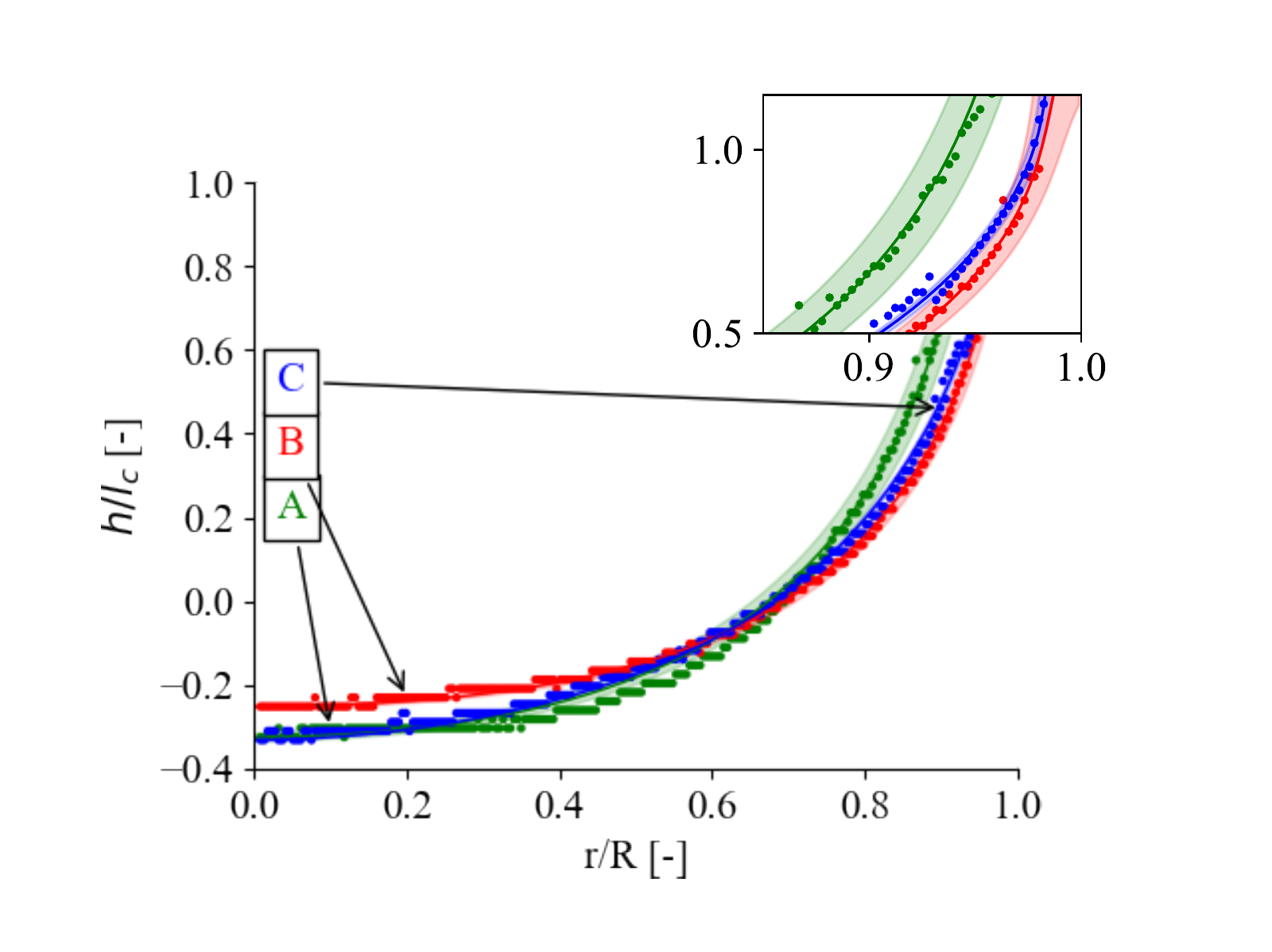}
    \caption{HFE7200 interface vs IC model}
    \label{fig:HFE inertia corrected}
    \end{subfigure}
    \hfill
    \hspace{-15mm}
    \begin{subfigure}[t]{0.49\textwidth}
    \includegraphics[width=8.5cm, trim={.8cm 1cm 0cm 1cm},  clip]{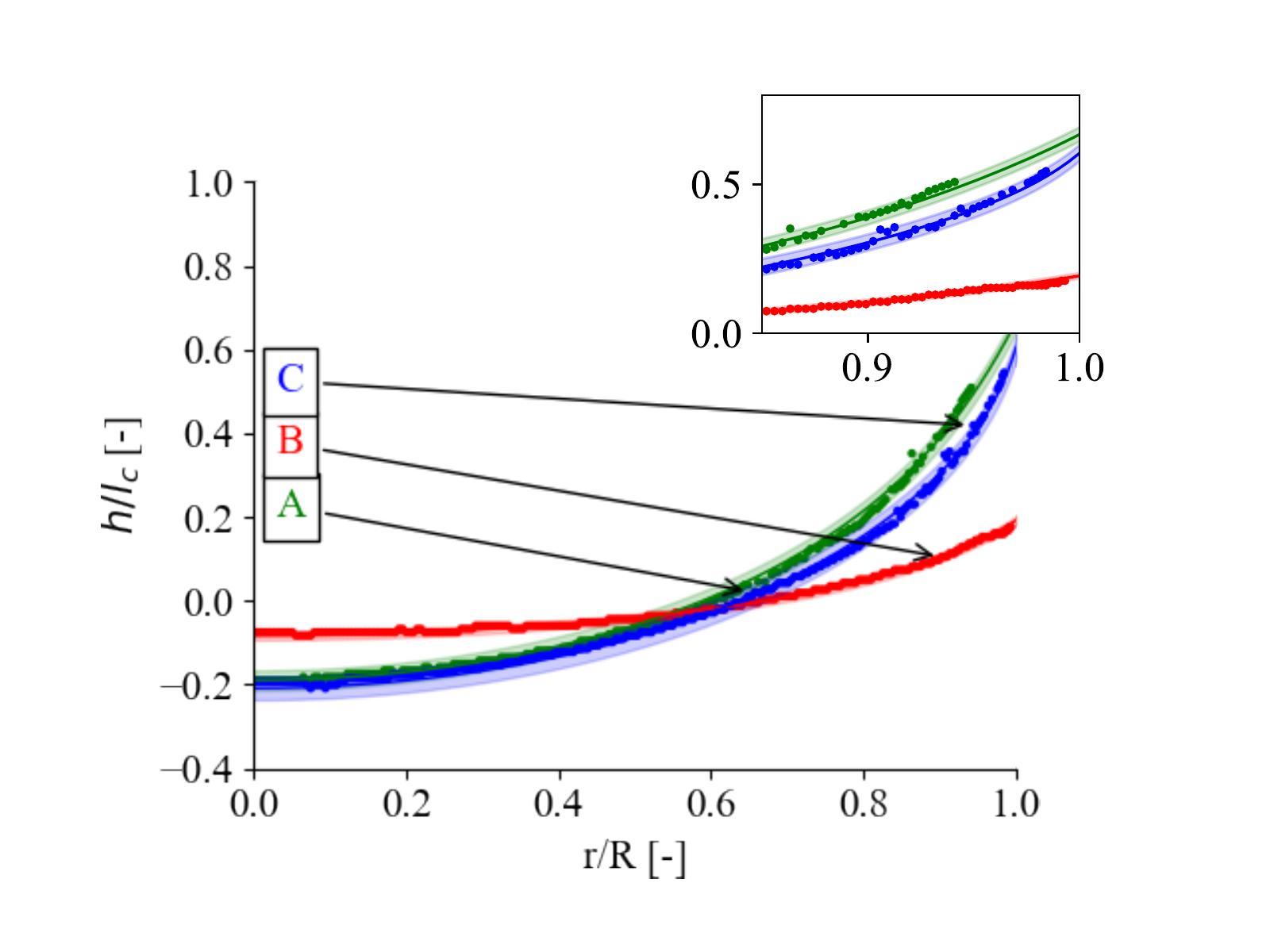}
    \caption{demineralized water interface vs IC model}
    \label{fig:WATER inertia corrected}
    \end{subfigure}
    \caption{Comparison of the experimental interface data (circular markers) with the corresponding model regression (solid lines) for three interface conditions (A-B-C). The plots on the left side correspond to an experiment with HFE7200; the plots on the right corresponds to an experiment with demineralized water. From top to bottom, each row shows the regression using the GB, the VG and the IC models respectively. The interface is shifted with respect to the mean value and normalized by the capillary length $l_c$. A zoom of the contact line region is shown on the top right each of each plot. At the largest acceleration and receeding conditions, both the GB and VG models overestimate the contact angle especially in the case of HFE7200.}
    \label{fig:modelRegression}
\end{figure*}

Here we analyze the experimental validation of the GB \eqref{eq:quasi-static}, VG \eqref{eq:quasi-steady} and  IC \eqref{eq:dynamic} models. The global performances of these models are shown in Figures \ref{fig:HFE regression error} and \ref{fig:WATER regression error} for a tests case with HFE7200 and demineralized water respectively. In these Figures, the upper plots show the time evolution of the objective function $\sqrt{J(\mathbf{w}^*)}$ used to solve the regression problem, where $J(\mathbf{w}^*)$ is evaluated at the optimal set of parameters $\mathbf{w}^*$, that is $\mathbf{w}^*=\theta$ for the GB and VG models and $\mathbf{w}^*=c_t,l_i,\theta$ for the IC model. $\sqrt{J(\mathbf{w}^*)}$ is the root mean square of the (optimal) Fréchèt distances \cite{Mendez2021_F} between the experimental points and the models' predictions. The bottom plots show the time evolution of the Weber number $We=\rho v_i^2(2R)/\sigma$, with $v_i=\dot{\bar{h}}$ the velocity of the spatially averaged interface as described in the previous section. 

Both figures should be analyzed together with figures \ref{fig:HFEexperiment} and \ref{fig:WATERexperiment}. In the case of HFE7200 (Figure \ref{fig:HFE regression error}), $We\approx 20$ at the first peak while it reaches at most $We\approx 4.5$ for demineralized water. During the first cycle ($t<0.2$s), both the GB and the VG models perform poorly --as expected-- since the interface experiences large acceleration and oscillations in this phase. For both models, the regression of the interface is harder in the case of HFE7200 than for water (leading to larger errors); we believe this is mostly due to the negligible contribution of surface tension in the first as compared to the second. Nevertheless, both GB \eqref{eq:quasi-static} and VG \eqref{eq:quasi-steady} models produce satisfactory results at largest time, as velocity and acceleration vanish.

The IC model \eqref{eq:dynamic} performs much better within the entire experiment. This is primarily due to the additional degrees of freedom in the parameter space. While for both GB and VG models oscillations of the interface in the center of the channel must induce changes in the contact angle (and vice-versa), this does not need to be the case for the IC. The additional term in \eqref{eq:pinertia} has the main merit of decoupling the interface dynamics far from the wall from the dynamics of the interface near the contact line.  

We illustrate the importance of such a correction by analyzing the interface shape in three conditions of largely different velocities and accelerations for the two fluids. These are indicated as A, B, C in Figures \ref{fig:HFE regression error} and Figure \ref{fig:WATER regression error}. In the case of HFE7200, the spatial-averaged interface velocity ($v_i=\dot{\bar{h}}$) and acceleration ($a_i=\ddot{\bar{h}}$) are
$(v_i,a_i)=$(-0.14 m/s, -1.6 m/s$^2$) in A, $(v_i,a_i)=$(0.1 m/s, 0.08
m/s$^2$) in B and $(v_i,a_i)$=(0.03 m/s, -0.01 m/s$^2$) in C. In the case of demineralized water, these are $(v_i,a_i)=$(-0.17 m/s, -1.23 m/s$^2$) in A, $(v_i,a_i)=$(0.03 m/s, -0.01
m/s$^2$) in B and $(v_i,a_i)=$(-0.04 m/s, -0.05 m/s$^2$) in C. \\

Figure \ref{fig:modelRegression} shows the comparison of the interfaces models for the three conditions A-B-C and for each of the fluids.  As expected, the GB (Figures \ref{fig:HFE gravity based}-\ref{fig:WATER gravity based}) and VG (Figure \ref{fig:HFE viscous based}-\ref{fig:WATER viscous based}) models are unable to describe the interface shape under strong acceleration (case A). Moreover, even at low acceleration levels it appears that despite the low regression error the GB model overestimates the slope of the interface near the contact line and thus the contact angle. This is more evident for the cases with HFE7200 (Figure \ref{fig:HFE gravity based}) than cases with water (Figure \ref{fig:WATER gravity based}), which had a higher contact angle.
 The VG model performs satisfactorily for $We<5$ (Figures \ref{fig:HFE viscous based}-\ref{fig:WATER viscous based}), while for higher $We$ (case B) this model also overestimates the slope of the interface near the contact line, especially in the case of HFE7200.
 
Figures \ref{fig:HFE inertia corrected}-\ref{fig:WATER inertia corrected} show the comparison of the same interfaces with the IC model. All the three conditions A-B-C are well described up to a close distance from the wall. 
In the case of HFE7200 (Figure \ref{fig:HFE inertia corrected}), the experimental interfaces differ at most $5\%$ of the capillary length $l_c$.

As pointed out in the previous section, the correction in the IC model is necessary to make the interface shape less dependant on the contact angle, and this is particularly important in receding conditions under large acceleration. The correction is effective in both fluids despite these feature largely different interface shapes. Particularly interesting is the test case B for water: this is taken at $t\approx 0.3$s, when the contact line is pinning at the wall while the rest of the interface is still advancing (and subject to a large inertia). 


\begin{figure*}
    \includegraphics[width=0.9\textwidth,trim={.0cm 0cm 1.1cm 0cm}]{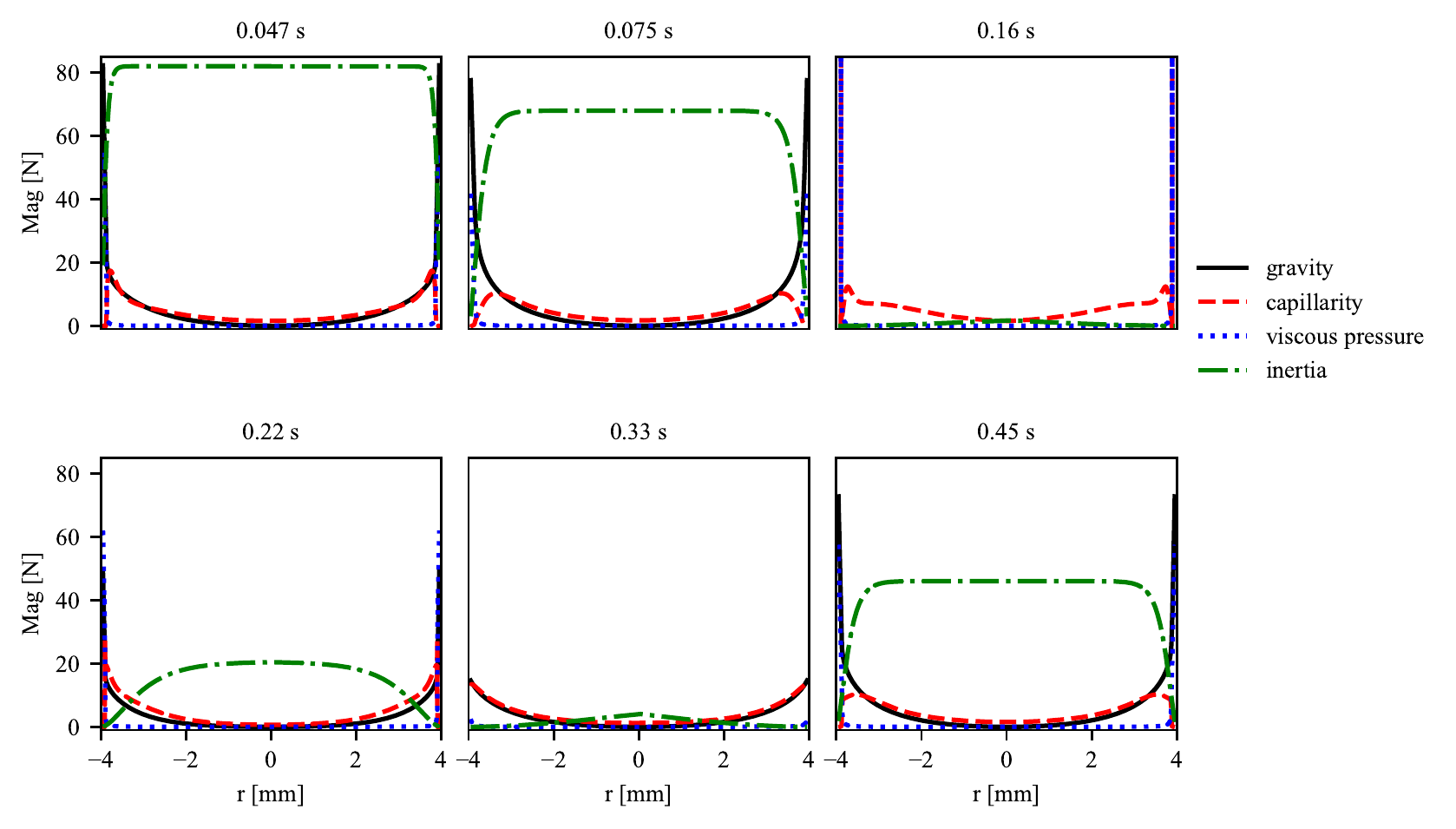}
    \caption{Profiles of the gravity (solid black line), capillary (dashed red line), viscous (dotted blue line), and inertia (dash-dot green line) terms across the channel in the IC model for different time of an experiment, for  the case of HFE7200. The inertial term is much larger than the others during the first moments of the experiment.}\label{fig:InterfaceBalanceHFE}
\end{figure*}

\begin{figure*}
    \includegraphics[width=0.9\textwidth,trim={.0cm 0cm 1.1cm 0cm}]{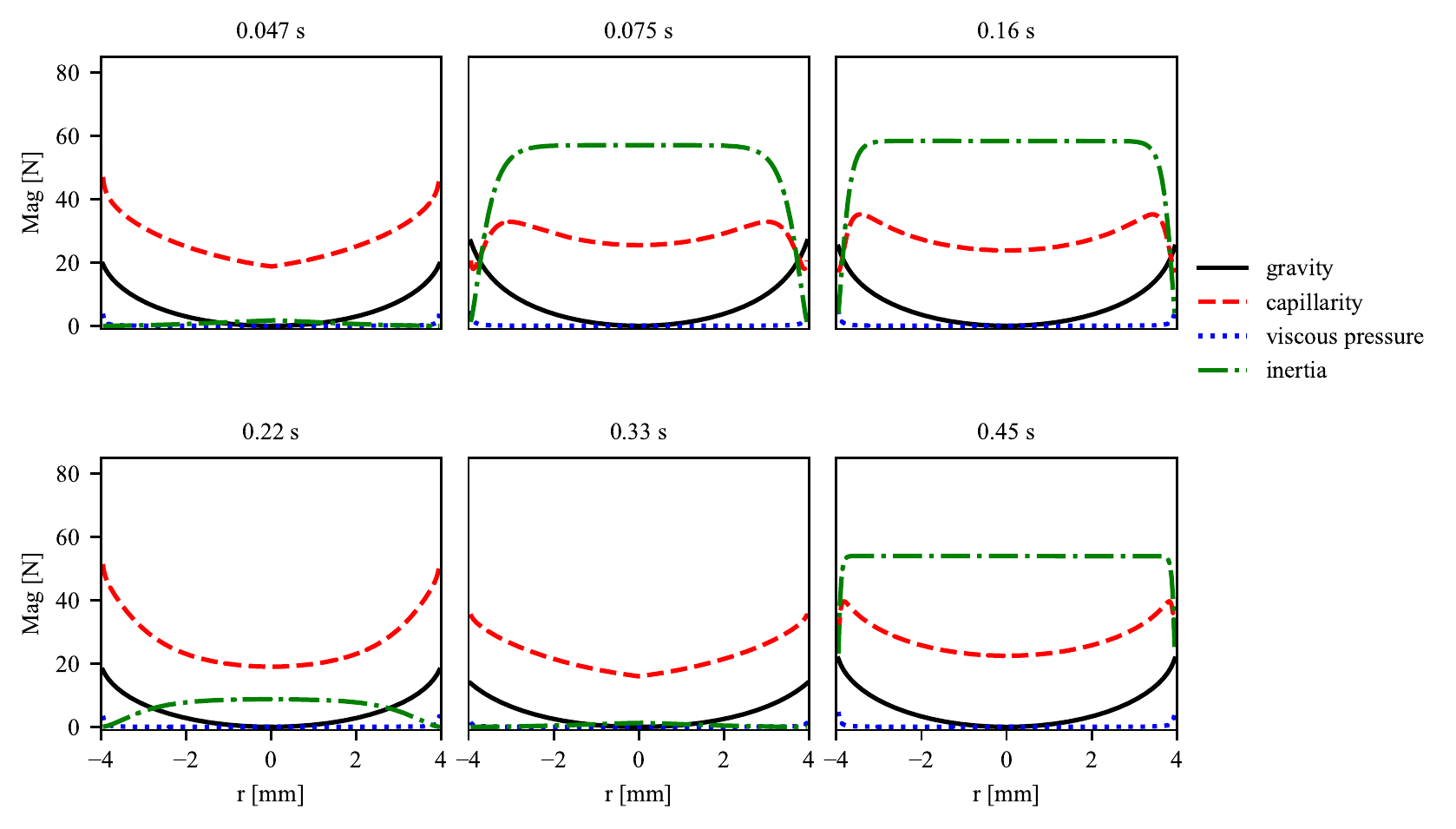}
    \caption{Same as Figure \ref{fig:InterfaceBalanceHFE} but with demineralized water.
    In this case the capillary term is much larger and comparable with the inertial term also in the first moments of the experiment.}\label{fig:InterfaceBalanceWATER}
\end{figure*}

Figures \ref{fig:InterfaceBalanceHFE} and \ref{fig:InterfaceBalanceWATER}  compare the magnitude of each term in the force balance of the IC model for both fluids. Each plot shows the distribution of the absolute value for each term as a function of the radial coordinate $r$. The title in each plot indicates the corresponding time (cf also Figures \ref{fig:HFEexperiment}-\ref{fig:WATERexperiment}).

These figures illustrate how different the force balance is in these two fluids. Yet, the inertial contribution plays a leading role in all the illustrated snapshots, especially far from the wall and in the case of HFE7200. In the case of water, the capillary force is the second most important contribution due to the larger surface tension, while in the case of HFE7200 this is the viscous term. This reaches large values near the contact line because of the small contact angle, while it is negligible in the case of water because of the much smaller interface slope at the wall.

Interestingly, the shape of the inertial correction is similar in the two cases, assuming a uniform value in the center of the channel at the largest acceleration and reaching the saturation at the distance at which the capillary term reaches a maximum (e.g. time 0.075 s and 0.16s in the case of water, 0.075 and 0.45s in the case of HFE7200).

\subsection{Transient contact angles}\label{subsec:result_angles}

\begin{figure*}
    \centering
    \begin{subfigure}[t]{0.47\textwidth}
    \includegraphics[width=8.5cm, trim={.2cm 0cm 0cm 1cm},  clip]{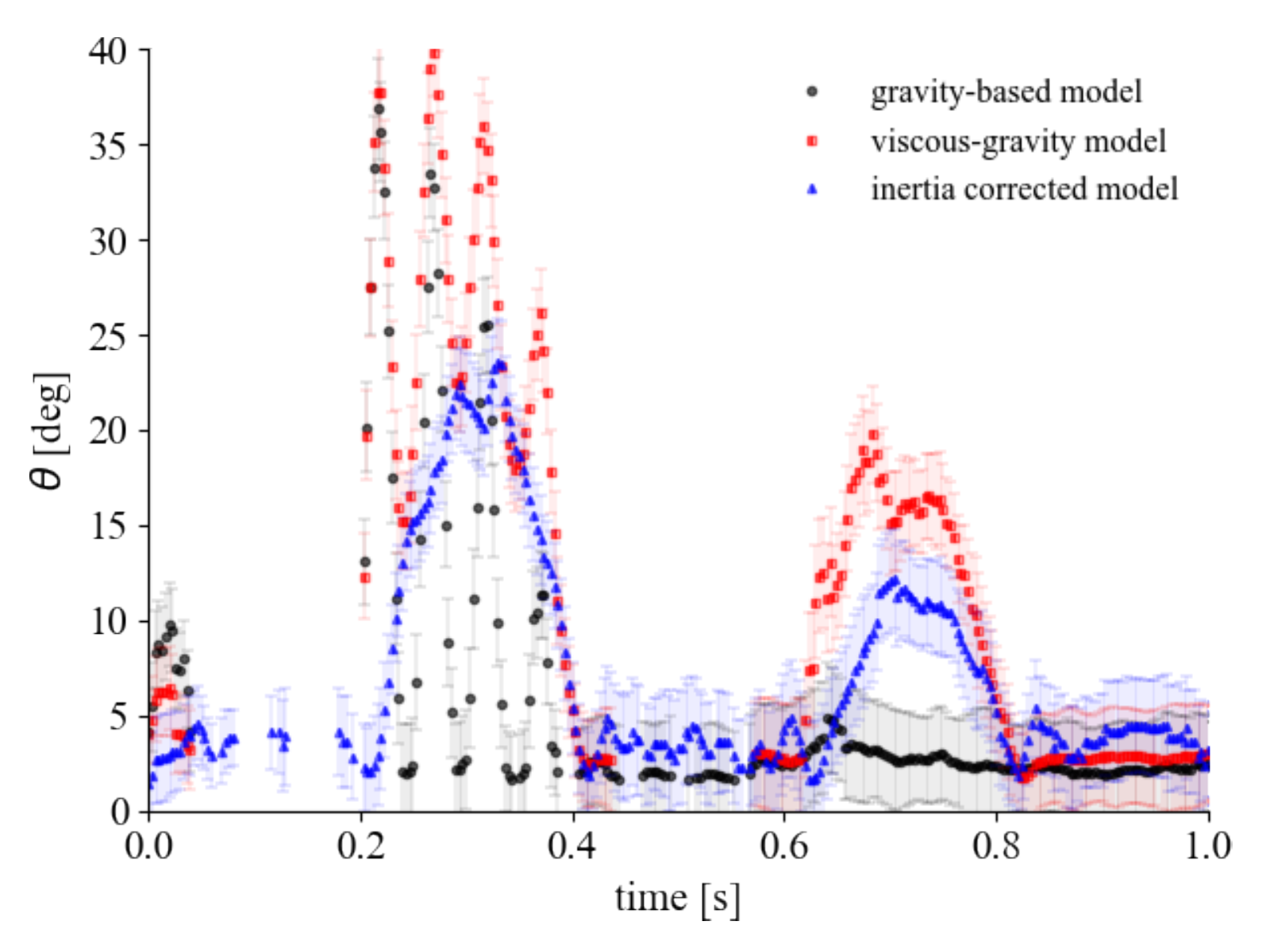}
    \caption{Contact angle evolution in HFE7200}
    \label{fig:HFE_CA_history} 
    \end{subfigure}
    \hfill
    \begin{subfigure}[t]{0.47\textwidth}
    \includegraphics[width=9cm, trim={1.6cm 1.6cm 3cm 1.2cm}, clip]{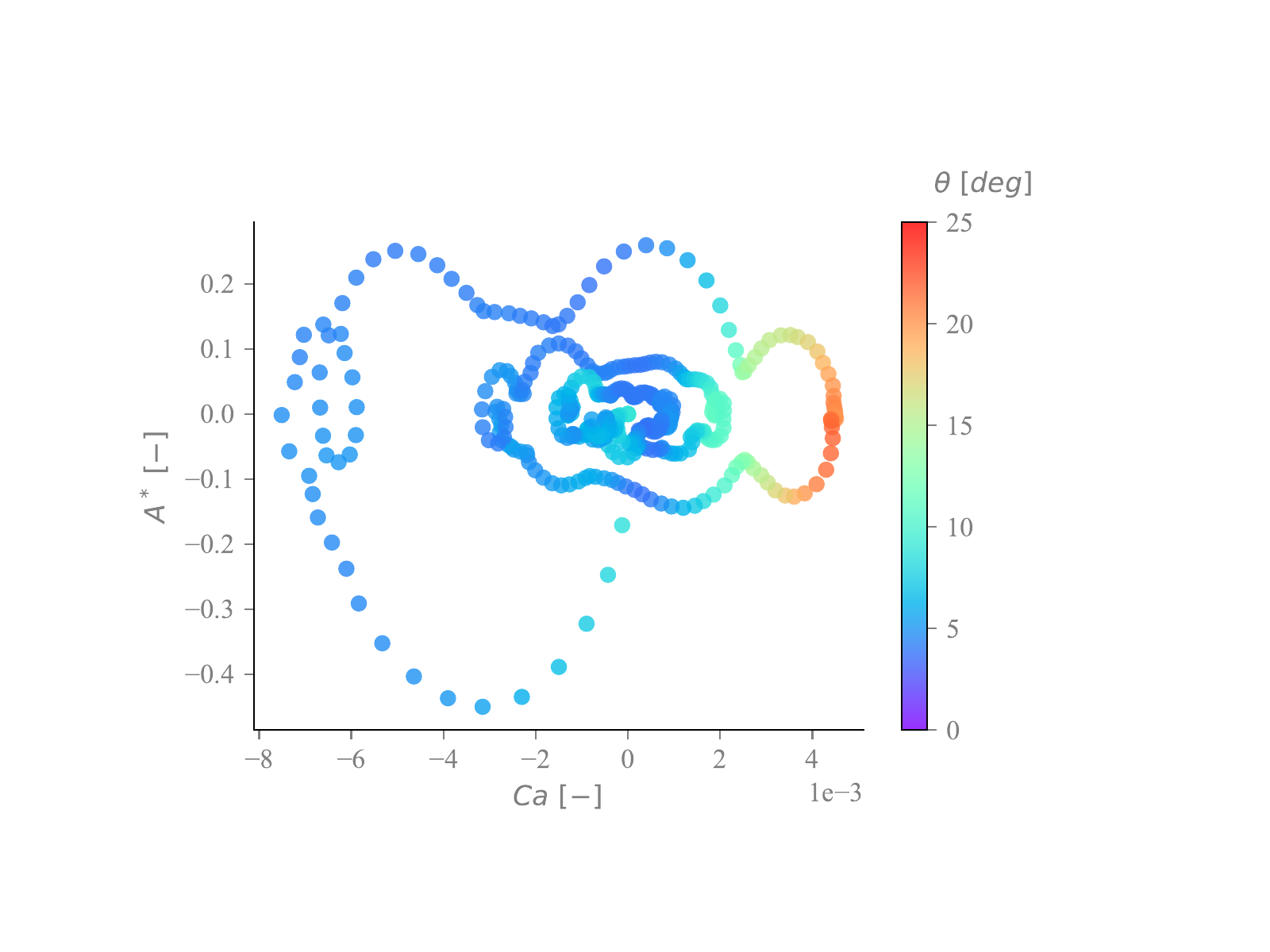}
    \caption{HFE7200 experiment}
    \label{fig:HFEOrbits} 
    \end{subfigure}
    \hfill
    \begin{subfigure}[t]{0.47\textwidth}
    \includegraphics[width=8.5cm, trim={.2cm 0cm 0cm 0cm},  clip]{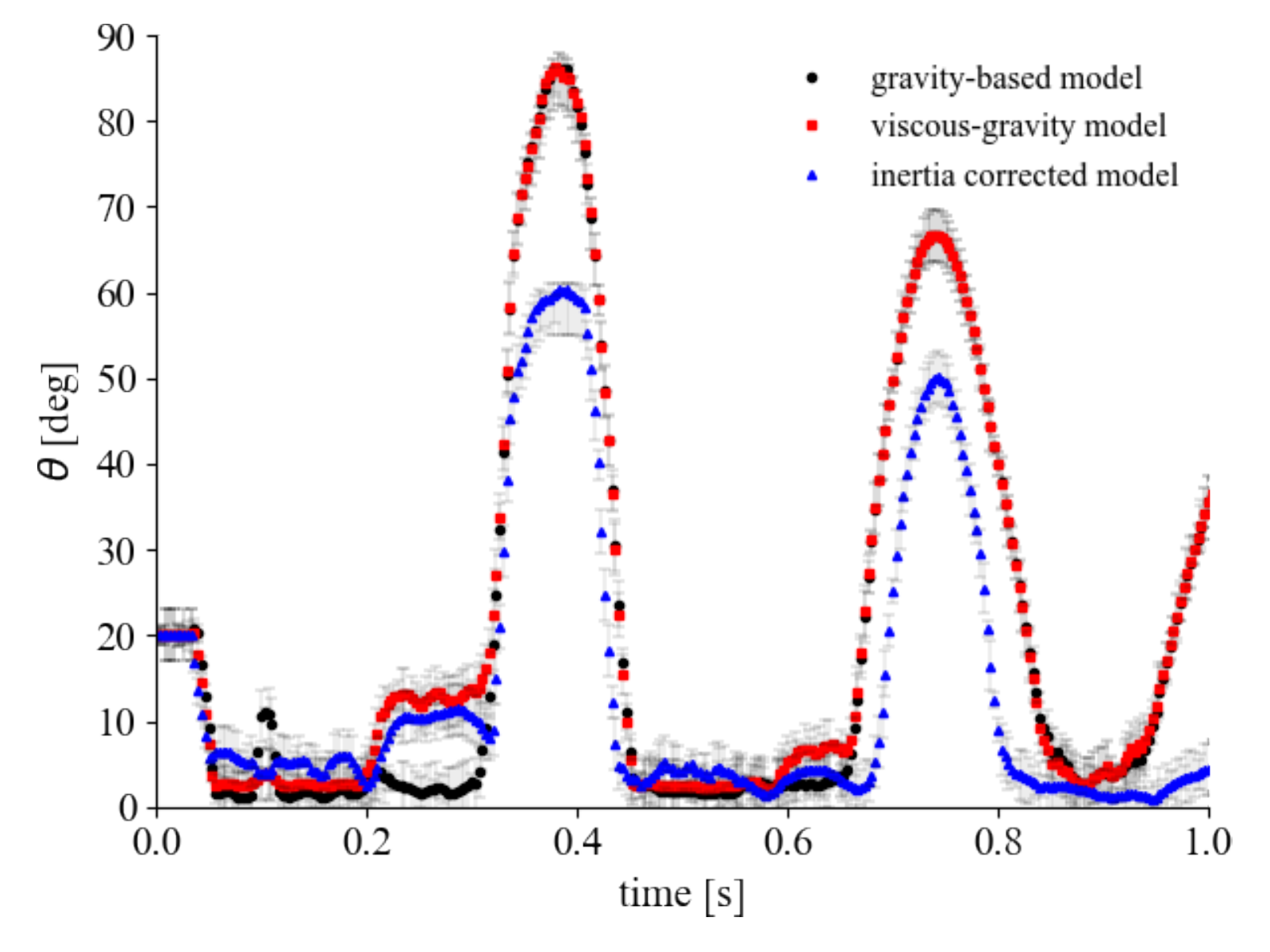}
    \caption{Contact angle evolution in water}
    \label{fig:WATER_CA_history} 
    \end{subfigure}  
    \hfill
    \begin{subfigure}[t]{0.47\textwidth}
    \includegraphics[width=9cm, trim={1.6cm 1.6cm 3cm 1.6cm},  clip]{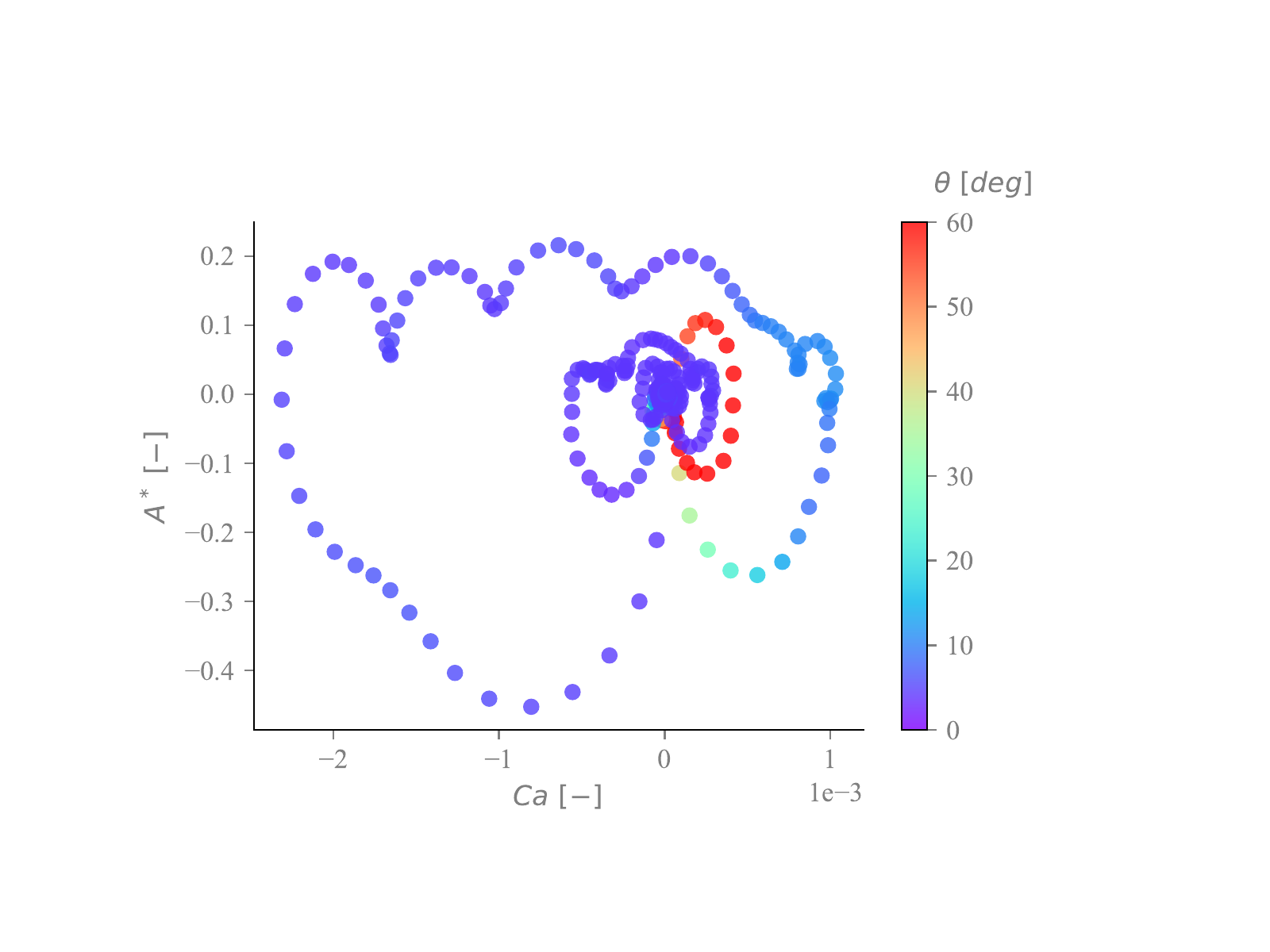}
    \caption{demineralized water experiment}
    \label{fig:WATEROrbits} 
    \end{subfigure}
    \caption{On the left: Evolution of the contact angle as a function of time according to  the gravity based (GB), viscous-gravity (VG) and inertial (IC) models for an experiment with HFE7200 (\ref{fig:HFE_CA_history}) and water (\ref{fig:WATER_CA_history}). All regressions with $\mathbf{\sqrt{J}}>25 \mu m$ are excluded. On the right: scatter plot mapping all the snapshots in the interface evolution to the contact line velocity (in terms of $Ca$) and acceleration (in terms of $A^*$). The markers are colored by the corresponding contact angle as predicted by the IC model and the values are reported in the colorbar on the right of each plot. }
\label{fig:CA_history}
\end{figure*}

Finally, we analyze the time evolution of the contact angle using the different interface models. Figure \ref{fig:HFE_CA_history} and \ref{fig:WATER_CA_history} show the contact angle as a function of time for a case with HFE7200 and demineralized water respectively. The data for each model contains only points for which the interface regression gives $\sqrt{\mathbf{J}}<25 \mu m$. Hence, fewer points are available for the GB and VG models in the first $\approx0.5s$.

In the case of HFE7200 (Figure \ref{fig:HFE_CA_history}), both GB and VG models predict large oscillations of the contact angle in the time interval $t\in [0.2,0.4]$s. This is the time interval between the first minima and the first maxima (see also Figure \ref{fig:HFEexperiment}) and is characterized by additional oscillations of the interface at the center of the channel (see also the oscillations in Figure \ref{fig:HFEacceleration}). These models must rely on the contact angle (see section \ref{sec:modeling}) to accommodate the interface oscillation and thus both models tie the interface dynamic to the contact angle evolution. But this link, in fact, is much weaker than what these models predict.

The IC model, leveraging the correction term, makes the contact angle independent from these inertial additional oscillations and predicts a smoother contact angle evolution. We consider this more reliable as the corresponding interface regression has a much lower error (see section \ref{subsec:result_interfaces}). Similar results are observed for the case of water (Figure \ref{fig:WATER_CA_history}), although the larger surface tension helps the GB and VG models. Nevertheless, as discussed in the previous section, these models significantly over-predict the contact angle.

We conclude the analysis by attempting to link the dynamic contact angle to the velocity and acceleration of the contact line. The plots on the right of figure \ref{fig:CA_history} map the dynamic contact angles measured by means of the IC model to the velocity and acceleration of the contact line for an experiment with HFE7200 (\ref{fig:HFEOrbits}) and demineralized water (\ref{fig:WATEROrbits}). The dimensionless contact line velocity is given by $Ca(t)=\dot{h}(R,t)\mu/\sigma$ while the dimensionless acceleration is scaled as $A^*(t)=\ddot{h}(R,t)/g$. Both $Ca$ and $A^*$ are given with their sign; hence $Ca<0$ corresponds to a receding interface while $Ca>0$ corresponds to an advancing interface.

\begin{figure}
    \centering
    \hspace*{-1cm} 
    \includegraphics[width=9cm,  clip]{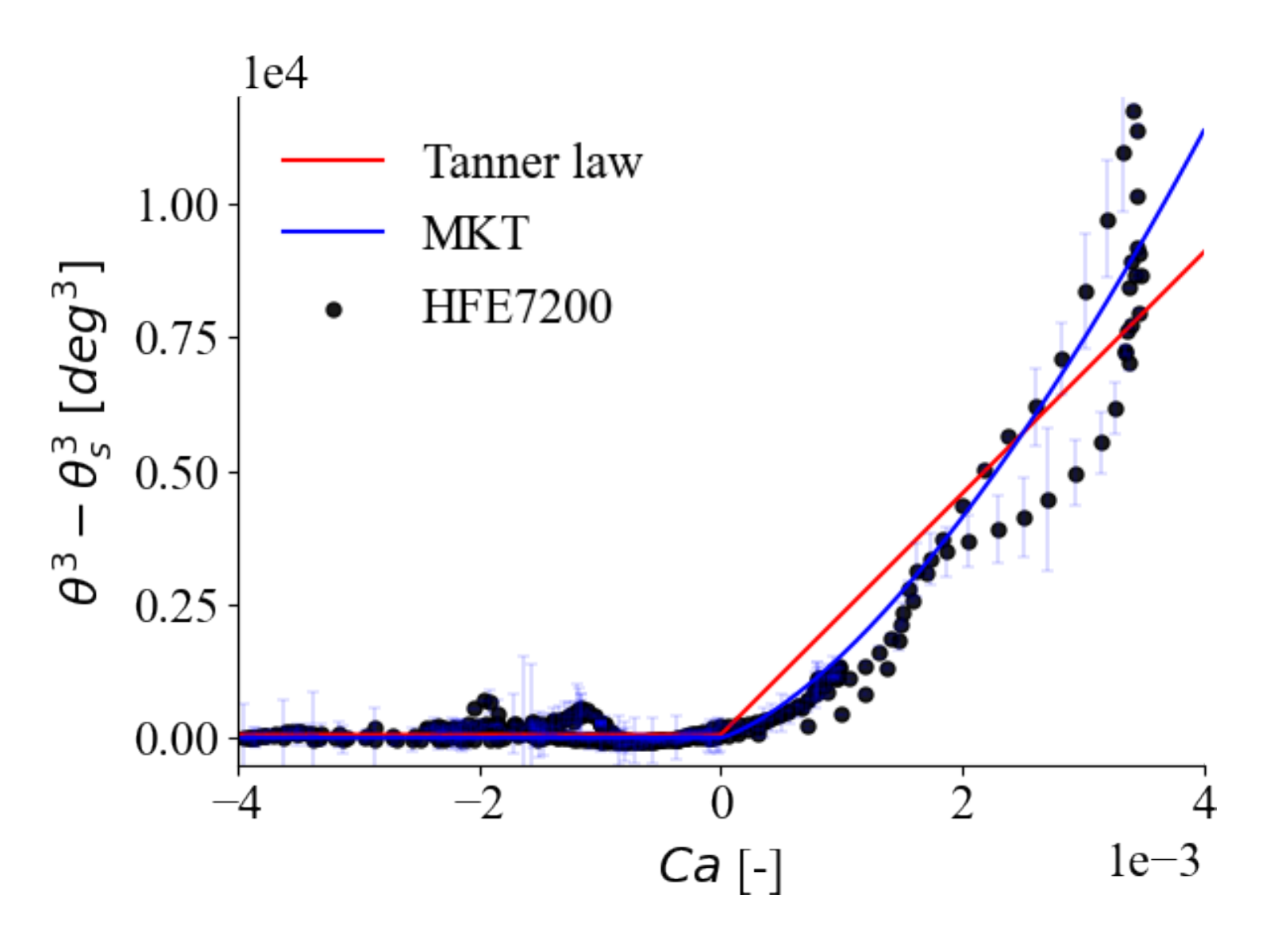}
    \caption{Comparison of dynamic contact angle data against capillary number for an experiment with HFE7200. In the case of receding contact line (negative $Ca$) the dynamic contact angle seemingly saturate at its static value. In the case of advancing contact line, the dynamic contact angle is compared with Tanner law (solid red line), and Molecular-Kinetic theory (solid blue line)}
    \label{fig:HFE_CAcorr}
\end{figure}

The marker are colored by the instantaneous dynamic contact angle $\theta(t)$ as predicted by the IC model. In the case of HFE7200, a clear trend is visible: the contact angle increases towards large $Ca$ regardless of $A^*$. At $Ca=0$ the contact angle $\theta$ varies between $4.5^{\circ}-5.5^{\circ}$ hence no appreciable wetting hysteresis is observed for HFE7200. This shows that classic contact angle correlations aiming at predicting $\theta$ as a function of $Ca$ can be identified. This is further illustrated in Figure \ref{fig:HFE_CAcorr}, which plots $\theta^3(t)-\theta^3_s$ as a function of the instantaneous capillary number $Ca(t)$ and compares the experimental data with the prediction of two well known correlations. These are the Tanner law  \cite{tanner1979spreading} and the molecular-kinetic theory (MKT) \cite{blake1969kinetics}.
The first express a linear trend in this plot with slope $\approx 12 Ca^{-1}$ while the second simplifies to $\theta\approx\cos^{-1}{(cos(\theta_s) -\sfrac{\zeta_w}{\mu} Ca)}$ with $\zeta_w= 12 \  mPa \ s$. The MKT law has been written as a function of the the capillary number and the \textit{contact line friction coefficient} $\zeta_w$   \cite{blake2006physics}. Clearly, given that the capillary number is of $\sim10^{-3}$ the  relationship between the capillary number and the dynamic contact angle is almost linear.

Although the relative uncertainties in this plot are considerable, due to the low contact angle, the trends suggest that such kind of correlations can be extended in the presence of significant acceleration and for the investigated system.

The same is not true for the case of demineralized water, as shown in Figure \ref{fig:WATEROrbits}. No clear trend can be identified, and the stick-slip motion might be only partially responsible for this since no relation can be found even at the largest $Ca$ (i.e. \emph{after} the pinning has occurred and the contact line motion re-established). The contact line pinning (leading to $Ca=0$) is observed whenever the contact angle approaches $\theta=60^o$. Whether this is the cause or the consequence of the pinning remains unknown. Nevertheless, despite the excellent accuracy of the IC model in the interface regression, no relation of the kind $\theta=f(Ca,A^*)$ can be identified since multiple contact angles are observed for the same velocity-acceleration conditions.

\section{Conclusion}\label{sec:conclusion}
This work analyzed the contact angle dynamics on low viscous liquids and in accelerating conditions, extending the literature on models and correlations for dynamic contact angles beyond the usual highly viscous liquids and steady contact line velocity. We consider two liquids: demineralized water (for its high surface tension) and HFE7200 (for its low contact angle).

Due to the high volatility of hydrofluoroethers and the low accuracy of direct techniques (e.g. TLM), previous studies only focused on the static contact angle \cite{sathyanarayana2012pool,cao2019electrophoretic} or assumed a constant zero value in dynamic conditions \cite{fries2008effect,fuhrmann2014heat}. To the authors' knowledge, this study shows for the first time the wetting dynamics of HFE7200.

We improve the dynamic contact angle characterization by using an indirect Meniscus Profile Method (MPM) based on a model of the gas-liquid interface. We extend previous approaches based on quasi-static (GB) \cite{petrov1993quasi} and quasi-steady (VG) \cite{maleki2007landau,iliev2011dynamic} interface models to the case of inertia-driven interface by proposing a heuristic `Inertia Corrected' (IC) model. The results show that the IC model performed well for both investigated liquids over the entire range of conditions. In contrast, comparing the GB and VG models with the experimental data showed that the coupling between the contact angle dynamic and the interface dynamics at the center of the channel is much weaker than GB and VG models predict.
The comparison of the dynamic contact angle against standard models (i.e. not accounting for the acceleration) showed that in the case of HFE7200 this could be linked to the instantaneous capillary number, as classic models would set, while for the case of demineralized water, this was impossible. An essential difference between the two configurations is that the contact line was found to pin multiple times in the case of water.

Future work will focus on analyzing whether the results obtained on HFE7200 hold for other highly wetting fluids and higher accelerations.

\begin{acknowledgments}
	D. Fiorini is supported by Fonds Wetenschappelijk Onderzoek (FWO), Project number 1S96120N. This work was supported by the ESA Contract No. 4000129315/19/NL/MG and by ArcelorMittal's `Minerva Contract'. Both partners are gratefully acknowledged. Finally, the authors thanks Mathieu Delsipee for its support and contribution in the preparation of the experimental set up.
\end{acknowledgments}
	
\section*{Data availability}
     The data that support the findings of this study are available on request from the corresponding author.

\section*{References}
\bibliographystyle{aapmrev4-1}
\bibliography{Main} 
	
\end{document}